\documentclass[aps,floats,twocolumn,epsf,prl,showpacs]{revtex4-1}
\usepackage{amsmath}
\usepackage{amssymb}
\usepackage{graphicx}
\usepackage[colorlinks=true,citecolor=blue,urlcolor=blue,linkcolor=blue,hyperfigures=true]{hyperref}
\usepackage{color}

\begin{document}

\title{Topology of SmB$_6$ determined by dynamical mean field theory}

\author{P. Thunstr\"om$^{1,2}$  and  K. Held$^1$}

\affiliation{$^1$Institute of Solid State Physics, TU Wien, 1040
Vienna, Austria}
\affiliation{$^2$Department of Physics and Astronomy, Materials Theory, Uppsala University, 751 20 Uppsala, Sweden}

\date{ \today }

\begin{abstract}
Whether SmB$_6$ is a topological insulator remains hotly debated. 
Our density functional theory plus dynamical mean field theory calculations are in excellent agreement with a large range of experiments, from the $4f^{5.5}$ intermediate valency to x-ray and photoemission spectra. Using the pole extended (PE) Hamiltonian, which fully captures the self-energy, we show that SmB$_6$ is a strongly correlated topological insulator, albeit not a Kondo insulator. The PE Hamiltonian is proved to be topological (in)equivalent to the ``topological Hamiltonian'' for (non-)local self-energies. The topological surface states are analyzed, addressing conflicting interpretations of photoemission data.

 \end{abstract}

\maketitle

The discovery of topology in the electronic band structure has added a whole new dimension to solid state physics\cite{RevModPhys.82.3045,RevModPhys.83.1057}. One of the most striking manifestation of a non-trivial topology is the emergence of robust metallic surface states in topological insulators. Such topological behavior has been established for the semiconductors mercury telluride \cite{Bernevig2006quantum} and bismut selenide \cite{nphys1270,nphys1274}, but the situation remains unclear for most other materials. For semiconductors, theory \cite{nature23268} is often ahead of experiment. That is, there are quite reliable predictions based on density functional theory (DFT)\cite{RevModPhys.61.689,perdew1992accurate} but no clear-cut experimental validation. Much more difficult, particularly for theory, are strongly correlated insulators for which the one-electron band picture breaks down. 

The archetype of such strongly correlated insulators is SmB$_6$, which was proposed by Dzero {\em et al.} \cite{PhysRevLett.104.106408} to be a topological Kondo insulator based on a (Kondo renormalized) non-interacting band structure and the topological Z$_2$ invariant as per \cite{ti:fu07}. 
However, the strong correlations and intermediate valency of SmB$_6$ give rise to an intricate multiplet structure \cite{Denlinger2013} with no clear adiabatic connection to a non-interacting system. Predictions based on a Kondo renormalization or the ``topological Hamiltonian'' \cite{PhysRevX.2.031008} of the system, constructed from the self-energy at zero frequency, may hence break down \cite{he2016topological}. In this letter we will instead focus on the less studied ``pole extended'' (PE) Hamiltonian \cite{savrasov2006many}, which fully captures the physical spectral function, and use it as a rigorous starting point for the topological classification \cite{Wang12}. 

On the experimental side, there is clear evidence of robust metallic surface states, such as a surface-dependent plateau in the low-temperature resistance \cite{PhysRevB.88.180405,Kim2013} and angular resolved photoemission spectroscopy (ARPES) data \cite{Jiang2013,Neupane2013,Xu13,hlawenka2018trivial} including its spin texture \cite{SpinSmB62013}. However, their topological origin has been questioned\cite{zhu2013polarity,hlawenka2018trivial}. Indeed, whether the low-energy electronic properties even originate from the surface or the bulk remains hotly debated. De Haas-van Alphen oscillations have been interpreted as stemming from both, the bulk \cite{Tan2015} and the surface \cite{li2014two,2016arXiv160107408D}. The same holds for the main ARPES features: \cite{PhysRevX.3.041024} vs.\ \cite{,Denlinger2013,2016arXiv160107408D}. The low-temperature linear specific heat has been shown to be predominantly a bulk effect \cite{PhysRevB.94.035127}; and the same holds for the optical conductivity within the band gap \cite{PhysRevB.94.165154}. On the other hand, it is was argued \cite{Fuhrman17} that these effects are not intrinsic but stem from $^{154}$Gd impurities which eludes a mass purification against  $^{154}$Sm. 

This controversy clearly calls for a better theoretical understanding. Because of the strong electronic correlations and well-localized $4f$ orbitals DFT +  Dynamical Mean Field Theory (DMFT) \cite{georges1996dynamical,kotliar2006electronic,held2007electronic} is the method of choice. 
There have been earlier DFT+DMFT \cite{2013arXiv1312.6637D,Kim2014,Peters2016}, DFT+Impurity \cite{thunstrom2009multiplet,Shick2015}, and DFT+Gutzwiller\cite{lu2013correlated} calculations which however did not capture the bulk band gap, the flatness of the $f$-bands, and the intermediate valence of SmB$_6$ all at the same time \cite{Denlinger2013}, due to various additional approximations. Indeed, the combination of these properties pose a hard theoretical challenge that requires an accurate many-body treatment of  the Sm $4f$ orbitals.

In this letter, we present charge self-consistent DFT+DMFT calculations for SmB$_6$. 
Our results provide an all encompassing picture of the bulk and surface properties of SmB$_6$, in excellent agreement with many experimental observations.
Analyzing the symmetry properties of the corresponding PE Hamiltonian \cite{savrasov2006many,Wang12} we show that SmB$_6$ is a strongly correlated topological insulator. 
Furthermore, we prove that the ``topological Hamiltonian'' \cite{PhysRevX.2.031008} is equivalent to the PE Hamiltonian if the self-energy is {\em local}, and pin-point the reason the former may fail for momentum-dependent self-energies.

{\em DFT+DMFT Method.}  
All calculations have been performed using the relativistic spin polarized toolkit (RSPt) \cite{Wills00,Grechnev07,thunstrom2012electronic,hitoshi2012high}, which is based on linearized muffin tin orbitals. This method allows the correlated Sm $4f$ orbitals to be readily identified and projected upon in the DMFT calculation \cite{Grechnev07}. The full Coulomb interaction, spin-orbit interaction, and local Hamiltonian for the Sm $4f$ orbitals are included in the DMFT impurity problem, which is solved using the exact diagonalization (ED) approach \cite{caffarel94exact,thunstrom2012electronic,liebsch2011temperature}. The calculations were performed at \mbox{$T = 100$~K} and iterated until charge self-consistency \cite{granas2012charge}. For additional details, such as the final bath state parameters and the double counting procedure \cite{solovyev1994corrected,anisimov1991band}, see the Supplemental Material (SM) \cite{SupplementalMaterial}.

{\em  Bulk.}  
The hybridization between the strongly localized Sm $4f$ orbitals and the surrounding orbitals is too weak to form a coherently screened (Kondo) ground state at 100 K. Instead we find an intermediate valence state with a thermal mixture of both Sm $4f^6$ and $4f^5$ configurations, with an average $4f$ occupation of $n_f\approx 5.5$, in good agreement with experiment \cite{mizumaki2009temperature}. The $4f^6$ contribution is predominately of $\Gamma_1$ character and the $4f^5$ of $\Gamma_8$ in agreement with non-resonant inelastic x-ray (NIXS) data \cite{Sundermann2018}. A detailed characterization of the thermal ground state is given in Table SII in the SM \cite{SupplementalMaterial}. 

\begin{figure}
\includegraphics[width=8.75cm]{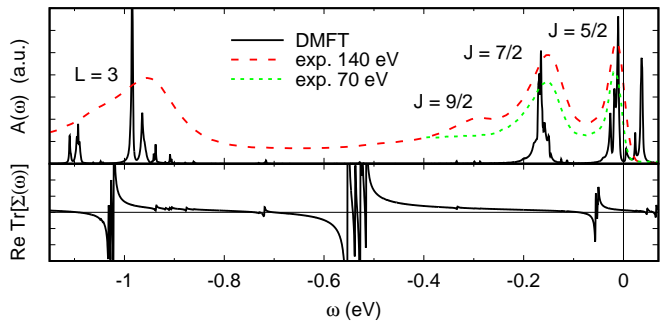}
\caption{Upper panel: Bulk $\mathbf k$-integrated DFT+DMFT spectrum of SmB$_6$ compared to experimental photoemission data\cite{Denlinger2013}. Sm $4f^6\rightarrow 4f^5$ transitions with the $J$ and $L$ quantum numbers of the final ($4f^5$) state are indicated.  Lower panel: The trace of the local self-energy (real part) shows distinct poles, but not within the bulk band gap.\label{fig:bulk}}
\end{figure}

The $\mathbf k$-integrated spectral function (DOS) reflects the intermediate valence and displays distinct $4f^6\rightarrow 4f^5$ and $4f^5\rightarrow 4f^4$ multiplet transitions, as shown in Fig.~\ref{fig:bulk} and the SM \cite{SupplementalMaterial}, respectively. The first peak below E$_F$ at -11~meV is to a $^6H_{5/2}$ $\Gamma_8$ final state \cite{SupplementalMaterial}. Upon closer inspection, we notice a $\Gamma_7$ subpeak with a maximum at -25~meV, too small to be well resolved in the experiment. The next peak around -170 meV corresponds to $J\!=\!7/2$ final states ($^6H_{7/2}$), and the on-resonance spectrum ($\hbar\nu =$ 140 eV) displays an additional $J\!=\!9/2$ peak around -300 meV.  The latter is hardly discernible in theory and at off-resonance ($\hbar\nu =$ 70~eV) as the direct transition is largely forbidden. 

These sharp multiplet peaks are generated from a many-body self-energy which has distinct poles, as seen in Fig.~\ref{fig:bulk} (lower panel). However, there is no pole within the narrow bulk band gap, as one would have for a Mott insulator. We will come back to the poles of the self-energy when discussing the topological properties.

\begin{figure}
\includegraphics[height=5cm]{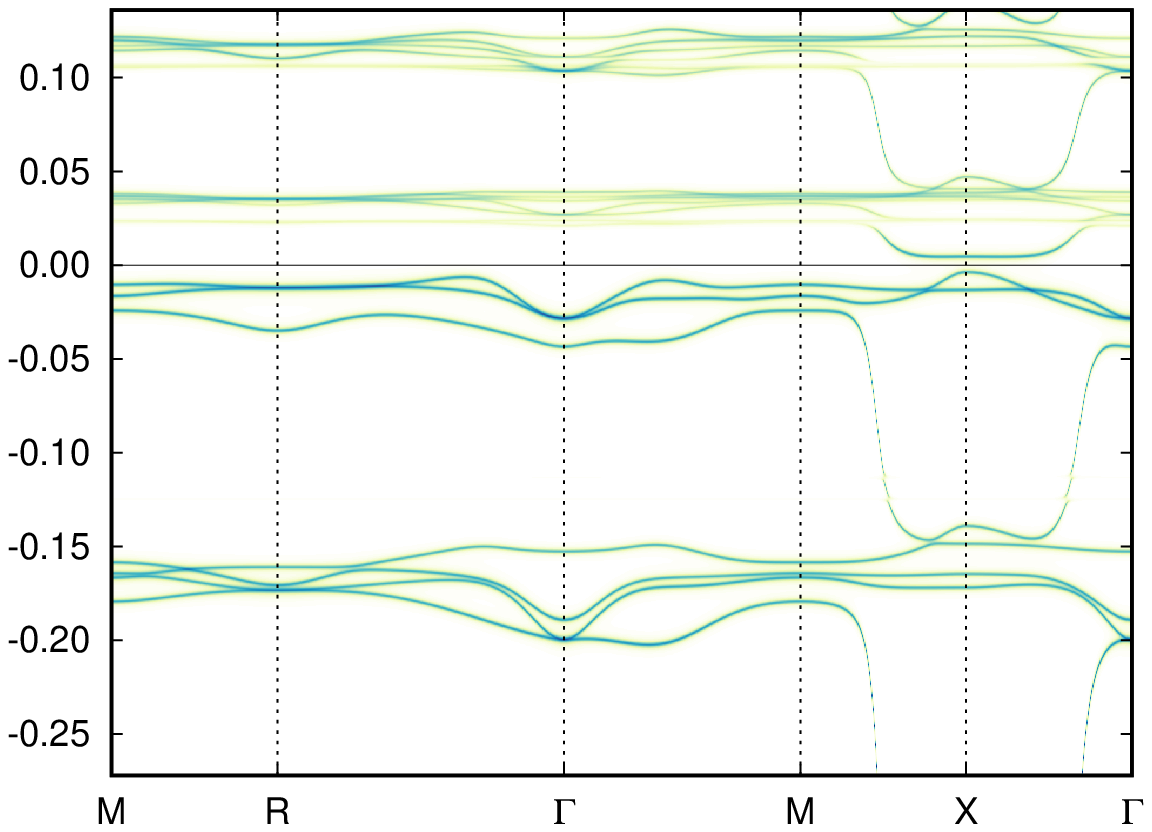}~\includegraphics[height=5cm]{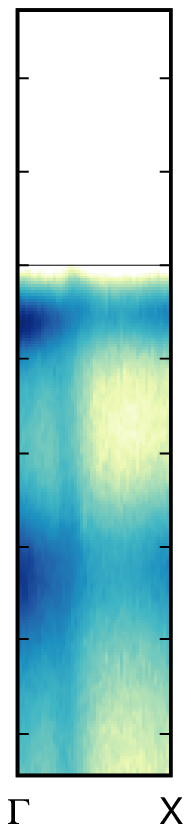}
\caption{Left: Bulk DFT+DMFT band structure of SmB$_6$ along the indicated high symmetry path through the Brillouin zone. Right: experimental ARPES data \cite{Denlinger2013} along $\Gamma -\mathrm{X}$. \label{fig:bulk-band}}
\end{figure}

Let us now turn to the momentum resolved spectral function in Fig.\ \ref{fig:bulk-band} (left). The Sm multiplet transitions discussed above result in flat $f$-bands each carrying a small fraction of the total weight. The Sm $5d$ orbitals on the other hand  hybridize strongly with the B $2p$ orbitals and form a parabolic band centered at the $X$-point. When this dispersive band crosses the flat $f$-bands close to the E$_F$ their hybridization leads to a small band gap of about 9 meV (16 meV peak-to-peak), close to the experimental value of $\sim$10 -- 20 meV \cite{bansil2016colloquium,Denlinger2013}. The spectral function agrees well with the bulk sensitive ARPES experiments of Denlinger {\em et al.} \cite{Denlinger2013} reproduced in Fig.~\ref{fig:bulk-band}.

Altogether, we find that the {\em local} DFT+DMFT self-energy gives an accurate description of the experimentally established bulk properties of SmB$_6$. Hence we can now turn to its topological properties with confidence.

{\em  Proof of non-trivial topology.}
The topological Z$_2$ invariant of Kane and Fu \cite{ti:fu07} can be determined, for an inversion-symmetric {\em non-interacting} system, from the parity ${\cal P}_{{\mathbf k}_i;m}$ of the (pairs of Kramers degenerate) bands $m$ below E$_F$ at the time reversal invariant moment (TRIM) momenta ${\mathbf k}_i$. If $\prod_{im} {\cal P}_{{\mathbf k}_i;m}=-1$, the system has a non-trivial topology. However, this procedure can not be directly applied to an interacting system, as the one-particle self-energy can split, smear out, and reduce the weights of the bands, and even make them fade away and reappear at shifted energies. Several suggestions of how to generalize the Z$_2$ invariant to the interacting case have been made\cite{PhysRevX.2.031008,Wang12}, but as common in the fast moving field of topological materials, the theory still needs to be developed in full.

The self-energy consists in general of a set of poles located on the real frequency axis \cite{gramsch2015lehmann,SupplementalMaterial}, clearly shown in Fig.~\ref{fig:bulk} (bottom). The local self-energy can hence be written as
\begin{equation}
\Sigma_{mn}(\nu) = \Sigma_{mn}(\infty) + \sum_l V_{ml}^\dagger \Big[ \nu - E_l\Big]^{-1} \!\! V_{ln}^{\phantom{\dagger}}.
\label{eqn:sigmapole}
\end{equation}
Here, $E_l$ and $V_{ml}^\dagger V_{ln}^{\phantom{\dagger}}$ are the pole position and weight, respectively; and $mn$ are the local spin-orbital indices. This pole structure of the self-energy is identical to that of a hybridization function 
where the physical orbitals $m$ hybridize via $V_{lm}$ with some {\em local} auxiliary orbitals $l$.
That is, we can make an exact mapping, akin to a purification of a mixed state, of the interacting bulk system with the
self-energy $\Sigma_{mn}$ to a non-interacting ``pole extended'' (PE) Hamiltonian with additional orbitals $l$ \cite{savrasov2006many,SupplementalMaterial}. As in the case of a purification, we recover the physical band structure by simply projecting the PE band structure onto the physical orbitals.
As this PE Hamiltonian is non-interacting, the topological invariant Z$_2$ of \cite{ti:fu07} can be straight forwardly applied \cite{Wang12}. If this Z$_2$ is non-trivial, the PE Hamiltonian has topological surface states, which manifest themselves in the physical system through the projection onto the physical orbitals. 

In the following, we show that SmB$_6$ is a topological insulator by  proving that (1) its PE Hamiltonian
is topologically non-trivial and that (2) the resulting topological surface states have a finite physical weight.

As for (1), the construction of the PE Hamiltonian requires the full pole structure of the self-energy, which can be difficult to obtain. Nevertheless, we can still assess its topological properties from some general considerations:

(i): As shown in Fig.~\ref{fig:bulk}, the self-energy has no pole at the Fermi energy. 

(ii): The DMFT self-energy is $\mathbf k$-independent in its local basis. The PE bands must therefore have a finite physical weight to be dispersive.

From (i) and (ii) it follows that since there is a finite band gap in the known physical spectral function, the PE Hamiltonian is insulating as well \footnote{A purely auxiliary band can not cross the band gap as this requires a finite dispersion.}. 

(iii): Only the correlated Sm $4f$ states carry a  DMFT self-energy, the Sm $5d$ orbitals and B $2p$ orbitals forming the dispersive bands have $\Sigma=0$. 

This is important since the auxiliary orbitals given by a local self-energy inherit the irreducible representation of the local correlated orbitals they derive from \footnote{The reason is simple, the symmetries of the system enforce that only local orbitals that share the same irreducible representation may hybridize.}. In our case, the auxiliary orbitals have the same  odd parity \footnote{when the inversion center is placed at the Sm atom} as the  Sm $4f$ orbitals. It is therefore enough to count the
pairs of the {\em non-interacting} bands of opposite (even) parity at the TRIM points to evaluate the Z$_2$ invariant!
These are 6, 6, 5, and 4  at $\Gamma$, $R$, $X$, and $M$, respectively \footnote{Excluding the semi-core and core states}.
The physical reason for the odd number at the $X$ point is that the dispersive
band of  Sm $5d$ and B $2p$ character is below E$_F$ at $X$. This  implies $\prod_{im} {\cal P}_{{\mathbf k}_i;m}=-1$, i.e., the PE Hamiltonian is topologically non-trivial.

This non-trivial topology is very robust against potential imprecisions in our DFT+DMFT calculation: they would need to shift an even parity orbital to the other side of E$_F$ at an odd number of TRIM points, which requires a shift by several eV, while maintaining the mixed valency. Furthermore, this shift must originate from the DFT potential and hence the electron density since $\Sigma=0$ for these orbitals.

As for (2): In general one expects that a non-trivial topology in the bulk implies robust metallic surface states. For a $\mathbf k$-dependent surface self-energy this need not be the case: In a gedanken experiment we can simply craft a  surface self-energy using auxiliary orbitals that mimics a large number of additional layers to the vacuum side.  Such a self-energy  makes the surface layer  indistinguishable from a bulk layer. Hence the topological surface states are completely absorbed by the surface self-energy. However, for a local DMFT self-energy, which is a good approximation for SmB$_6$, a complete absorption is not possible. The local self-energy may vary from layer-to-layer (perpendicular to the surface), but it still fulfills (ii). Hence, the dispersive topological surface states must carry a finite physical weight as they cross the bulk band gap.

Let us also address the concerns of a possible break down \cite{he2016topological} of the ``topological Hamiltonian'' \cite{PhysRevX.2.031008}. For a ${\mathbf k}$-independent self-energy such a breakdown is not possible as long as it is defined, i.e., (i) $\Sigma(0) < \infty$. 
We show  this in the  SM \cite{SupplementalMaterial} by adiabatically detaching the auxiliary orbitals from the PE Hamiltonian. The resulting Hamiltonian is identical to the ``topological Hamiltonian'', apart from the detached auxiliary orbitals, which due to their purely local character do not contribute to the non-trivial topology. Hence, the ``topological Hamiltonian'' and the  PE Hamiltonian are topologically equivalent when the self-energy is local. For a ${\mathbf k}$-dependent self-energy, on the other hand, the auxiliary orbitals can contribute in a non-trivial way to the elementary band representations \cite{nature23268,SupplementalMaterial}. This can manifest itself as additional topological surface states that gap out the topological surface states predicted by the topological Hamiltonian, which is consistent with the break down seen in \cite{he2016topological}.

\begin{figure}
\parbox[b]{7.1cm}{
\includegraphics[width=7cm]{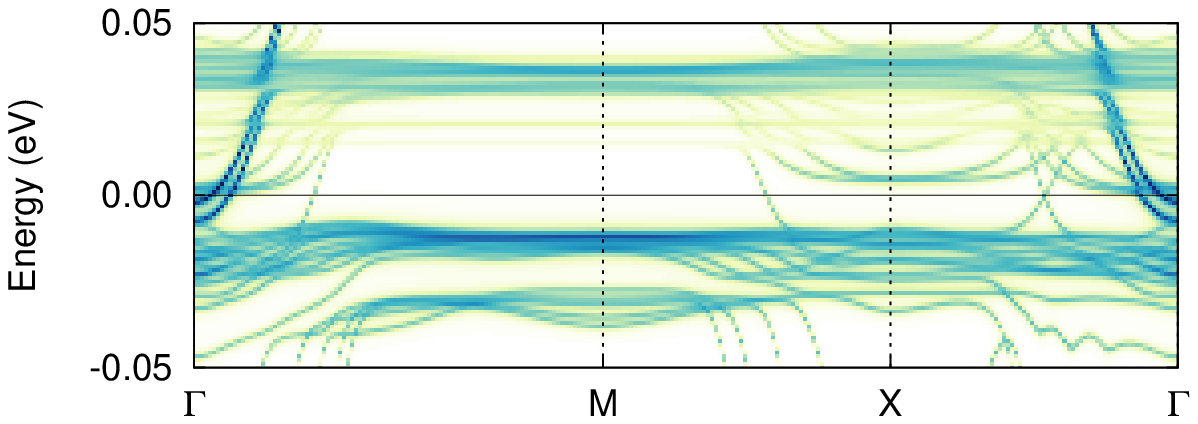}\\
\includegraphics[width=7cm]{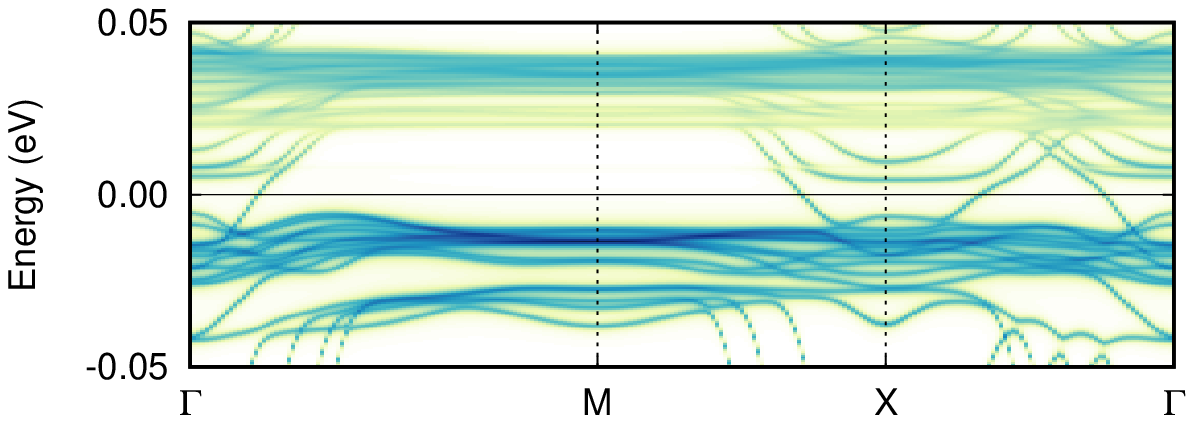}
}\hspace{2mm}\includegraphics[width=1.15cm]{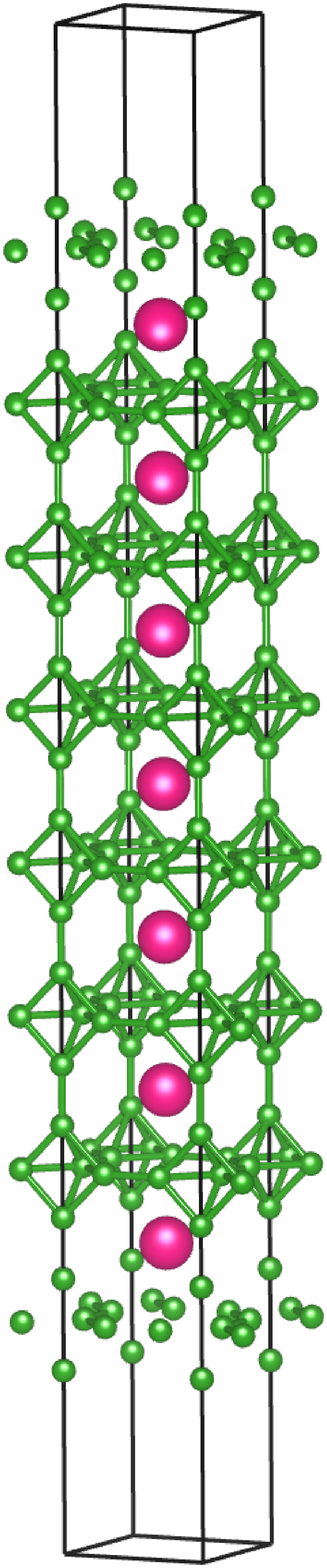}
\caption{DFT+DMFT spectrum resolved for $k_x$ and $k_y$ of a periodic SmB$_6$ supercell with (top left) and without (bottom left) B$_6$ surface termination. The former is depicted on the right. Both terminations show topological surface states that cross the bulk band gap. The B$_6$ termination has additional trivial surface bands associated with the B$_6$ dangling bond.\label{fig:slab-band}}
\end{figure}

{\em  Surface states.}
To address the conflicting interpretations of the experimental photoemission data\cite{Denlinger2013,Xu13,hlawenka2018trivial}, we performed additional DFT+DMFT slab calculations. 
The [001] SmB$_6$ supercell, shown to the right in Fig.~\ref{fig:slab-band}, was used with two different terminations to represent the surface patches of sputtered films\cite{zabolotnyy2018chemical} and cleaved samples\cite{hlawenka2018trivial}. The former has a terminating B$_6$ layer (depicted without bonds) and the outermost Sm atom is Sm$^{3+}$, while the latter lacks the B$_6$ layer and is instead terminated by Sm$^{2+}$ surface atoms. The B$_6$ termination has trivial surface bands associated with a B$_6$ dangling bond \cite{zhu2013polarity} that disappears when the surface reconstructs \cite{zhu2013polarity,hlawenka2018trivial,Denlinger2013}. To effectively mimic this partial surface reconstruction in our computationally expensive DMFT calculation, we simply apply an additional 0.5 eV potential to the 2p orbitals of the outermost B atom in the final step. This potential shifts the trivial surface band, but does not directly affect the topological surface states which live, as we will see, in the subsurface layer.

The resulting DFT+DMFT spectra are presented in Fig.~\ref{fig:slab-band} (left). We recover the flat Sm $4f$ bands of the bulk at -14meV and +40meV as well as the $m_J=\pm 1/2$, $J=5/2$ band directly above E$_F$ at the $X$ point. On top of these bulk states surface bands emerge within the bulk band gap. These metallic surface bands are mainly associated with the subsurface Sm layer, as the Sm $4f$ states of the outermost Sm layer are shifted away from E$_F$ due to their pure Sm$^{2+}$ or Sm$^{3+}$ characters. In the SM \cite{SupplementalMaterial} we further confirm the topological protection of the surface bands by applying artificial potentials to the subsurface atoms. The topological surface states simply shifts deeper into the material when an additional (time-reversal symmetric) perturbing potential is applied to the (sub)surface layer. On the contrary, an out-of-plane magnetic field, which breaks the time-reversal symmetry, makes the surface bands detach from each other and form a band gap\cite{SupplementalMaterial}.

\begin{figure}
\noindent\begin{tabular}{p{0.2cm} p{3.2cm} p{3.2cm} p{1.6cm}}
& \centering B$_6$ termination & \centering Sm termination & \centering Exp.
\end{tabular}
\rotatebox{90}{\parbox{3.3cm}{\centering E$_\mathrm{F}$}}\hspace{0.2mm}\includegraphics[width=3.3cm]{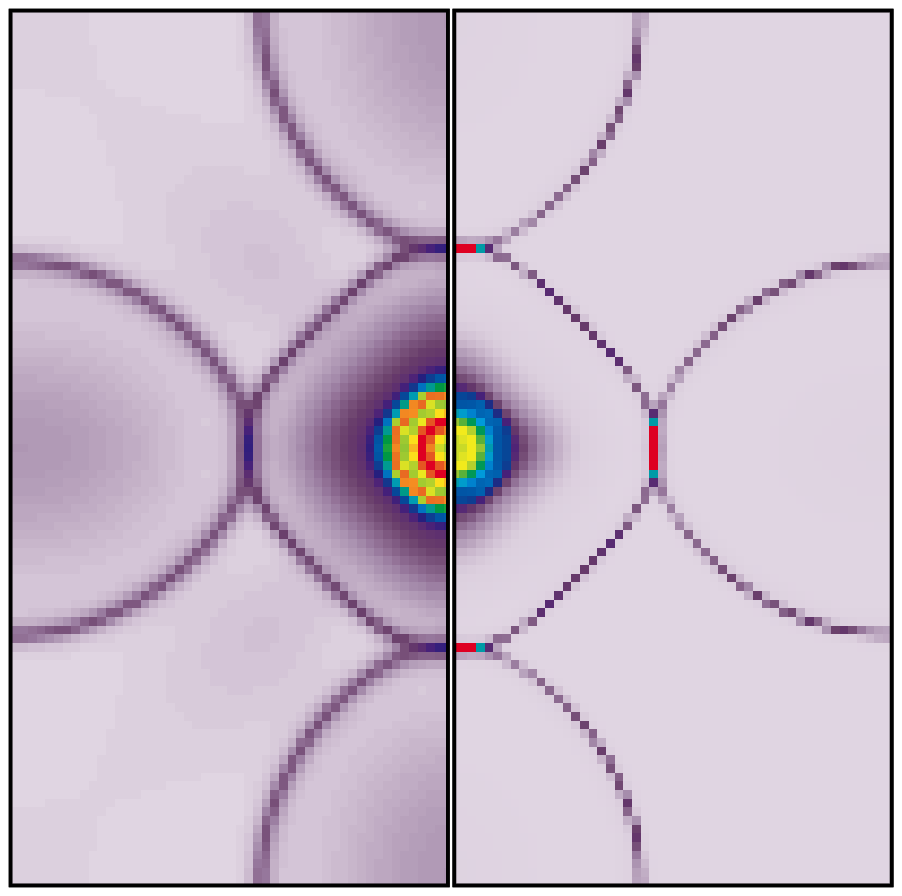}\hspace{0.5mm}\includegraphics[width=3.3cm]{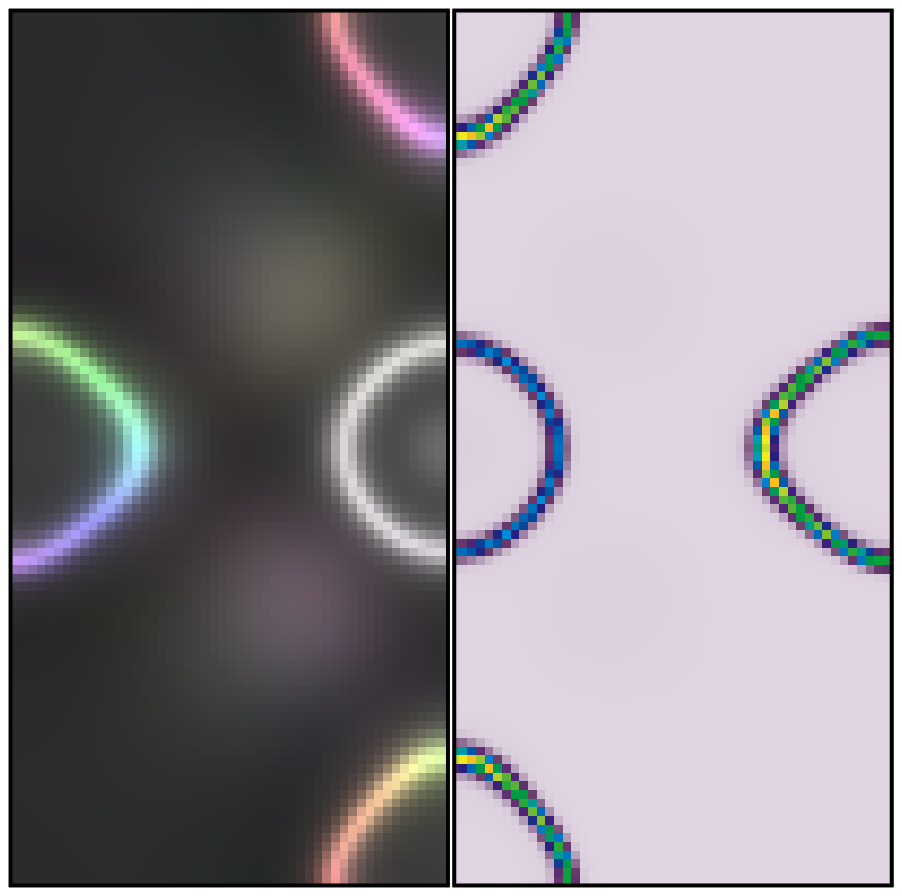}\hspace{0.5mm}\includegraphics[width=1.65cm]{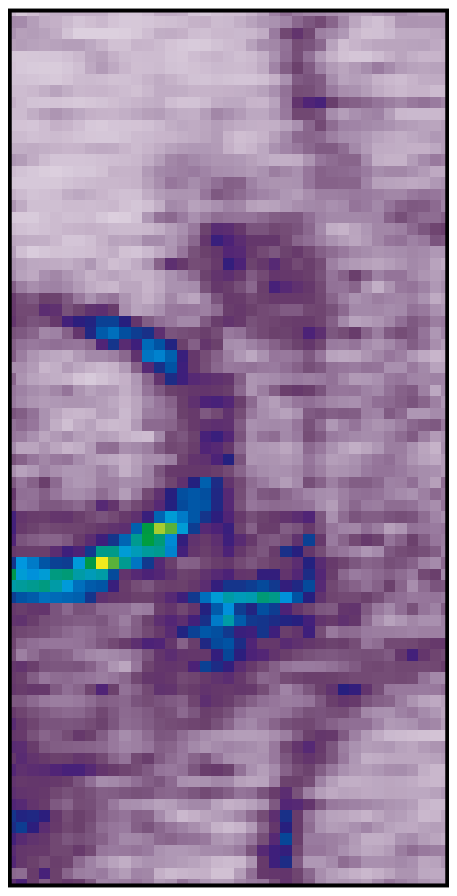}\\
\rotatebox{90}{\parbox{3.3cm}{\centering -5 meV}}\hspace{0.6mm}\includegraphics[width=3.3cm]{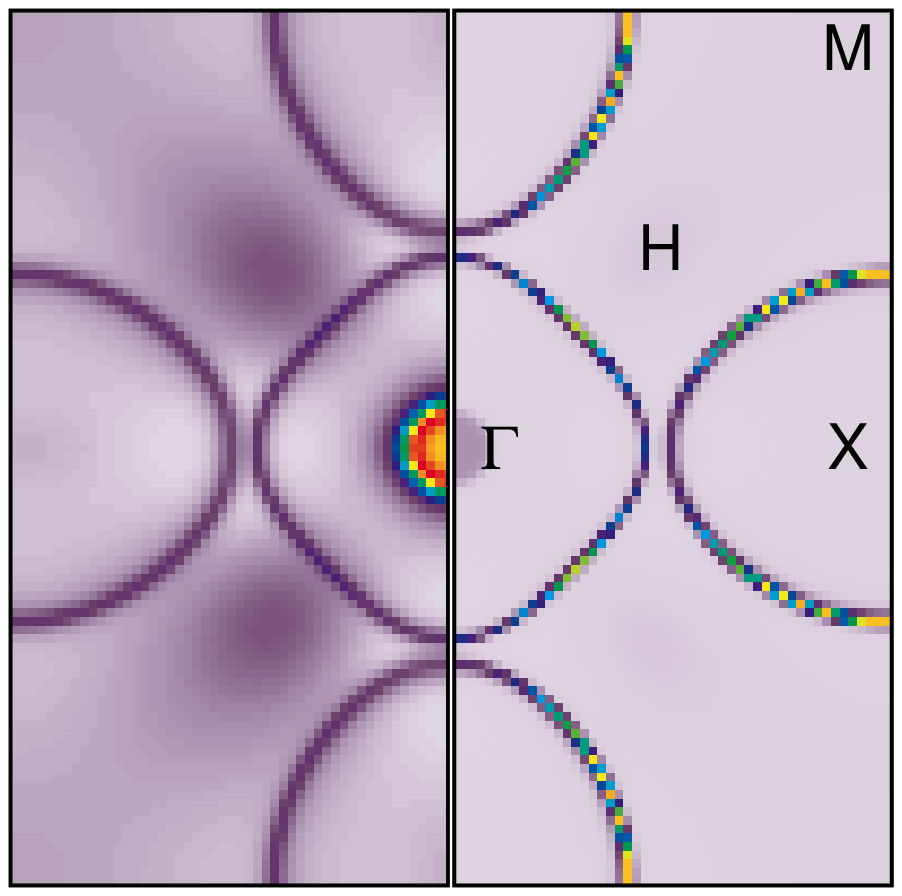}\hspace{0.5mm}\includegraphics[width=3.3cm]{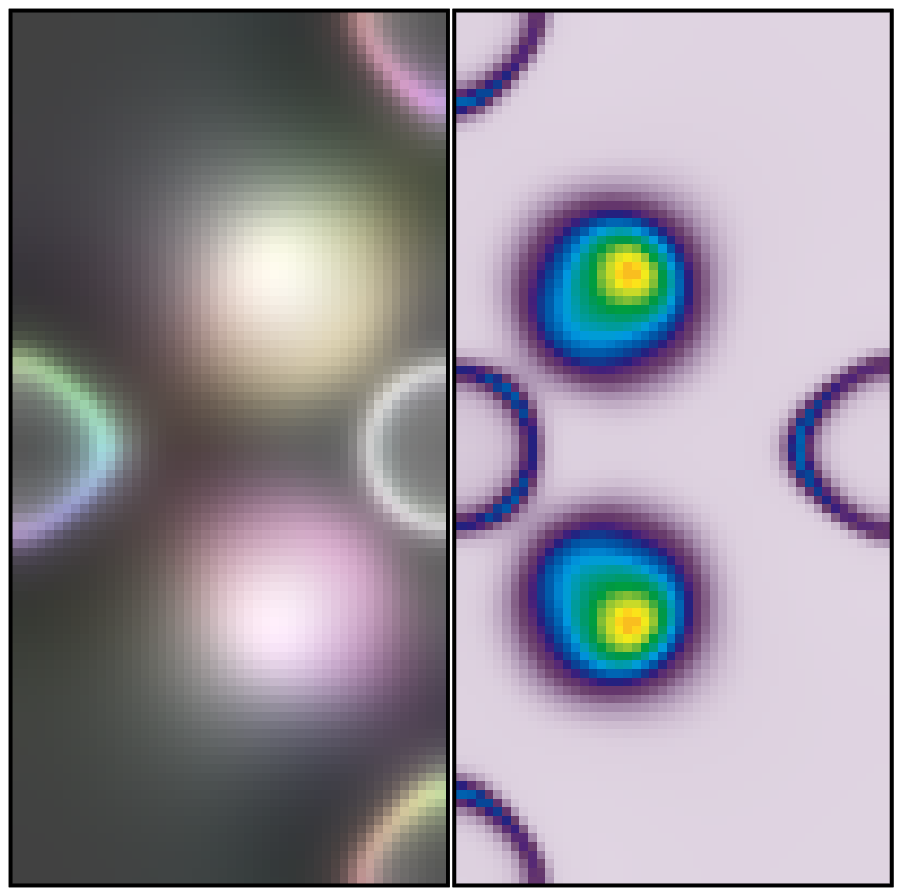}\hspace{0.5mm}\includegraphics[width=1.65cm]{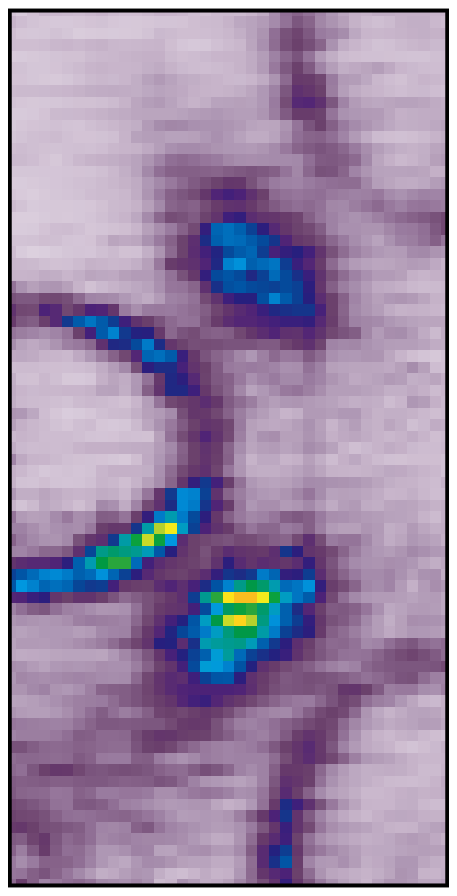}\\
\noindent\begin{tabular}{p{0.2cm} p{1.5cm} p{1.5cm} p{1.5cm} p{1.5cm} p{1.6cm}}
& \centering Total & \centering Sm (ss) & \centering $\langle \vec{S} \rangle$ (ss) & \centering Sm (ss) & \centering Total
\end{tabular}
\caption{Surface states resolved for $k_x$ and $k_y$ at E$_F$ (top) and -5 meV below E$_F$ (bottom). Left: B$_6$ termination. Middle: Sm termination. Right: experiment \cite{Denlinger2013}. The B$_6$ termination shows the total and Sm 4f subsurface (ss) contribution. The Sm termination shows the Sm 4f (ss) contribution and its spin projection $\langle \vec{S} \rangle$ mapped on the color wheel. The maximum total intensity for the B$_6$ termination (left most panel) is approximately 7 times larger than its Sm (ss) projection.\label{fig:surface}}
\end{figure}

Fig.~\ref{fig:surface} shows cuts of the spectral function at E$_F$ and -5~meV below. 
The surface states form large and intermediate sized pockets around both the $X$ and $\Gamma$ points for the B$_6$ and Sm termination, respectively.
The $X$ pockets show a strong spin polarization in agreement with spin-resolved ARPES experiments \cite{Xu13,SupplementalMaterial}. 
Ref.~\cite{hlawenka2018trivial} reports both the large $X$ and $\Gamma$ pockets for the B$_6$ terminated surface, although they interpret the large $\Gamma$ pocket as an umklapp state. The trivial ``Rachba split'' surface states reported close to the $\Gamma$ point directly after cleaving \cite{hlawenka2018trivial} seem instead to correspond to the shifted B$_6$ derived trivial surface bands. Ref.~\cite{Denlinger2013} shows the intermediate sized $X$ pockets and the H features at -5 meV associated with the Sm terminated surface, reproduced in Fig.~\ref{fig:surface}, as does \cite{Jiang2013,Xu13}, but the $\Gamma$ pocket is seemingly missing. However, on closer inspection of the data, in particular in the second Brillouin zone, there is clear evidence of a matching surface derived $\Gamma$ pocket, which again has been interpreted as an umklapp state \cite{Jiang2013,Xu13}. Hence, to finally settle the apparent discrepancy between theory and experiment, we suggest a critical experimental reexamination of this umklapp assignment. For example, our data for the Sm termination suggest that the $\Gamma$ pocket has a much weaker spin-polarization than the $X$ pocket.

{\em Conclusion.}
The appropriateness of DFT+DMFT for describing the strong correlations in SmB$_6$ and the excellent description of various experimental properties gives us confidence that we have achieved an accurate theoretical description of SmB$_6$. 
We determined the topological nature in a rigorous way from the symmetry properties of the PE Hamiltonian, formed by mapping the pole structure of the self-energy onto a set of auxiliary orbitals \cite{savrasov2006many,Wang12}. We made the key observation that also the auxiliary orbitals contribute to the ``elementary band representations'' \cite{nature23268} of the PE Hamiltonian, which clarifies the fundamental topological role of the poles of the self-energy. Nevertheless, we prove that for a local DMFT self-energy it is enough to count the bands of the opposite parity channel at the TRIM points, under the condition that the self-energy does not have a pole at E$_F$. This band analysis shows that SmB$_6$ constitutes a topological insulator. 

In general, if the positions of the bands in one parity channel is well-described by DFT, then the addition of a {\em local} self-energy in the other parity channel will not change the topology of the system as long as it remains insulating. If the local self-energy has poles within the bulk band gap topological surface bands will still appear in the spectral function, but they may gradually flatten out and vanish as they approach the poles.

We have also shown, through a simple qualitative argument, that a {\em non-local} symmetry preserving surface self-energy can completely absorb any topological surface bands. However, our data suggests that the ``missing'' $\Gamma$ pocket in SmB$_6$ instead is simply disguised as umklapp states \cite{hlawenka2018trivial,Jiang2013,Xu13}.

\begin{acknowledgments}
We are highly indebted  to J.~D.~Denlinger and J.~W.~Allen for making available their ARPES data. We would also like thank Annica Black-Schaffer and Dushko Kuzmanovski for useful discussions. This work has been  financially supported in part  by European Research
Council under the European Union's Seventh Framework Program
(FP/2007-2013) through ERC grant agreement n.\ 306447. The computations were performed in part on resources provided by SNIC, under project SNIC 2018/3-489.
\end{acknowledgments}

\bibliographystyle{apsrev4-1}
\bibliography{full2}
\clearpage
\pagebreak
\onecolumngrid
\renewcommand{\thefigure}{S.\arabic{figure}}
\renewcommand{\theequation}{S.\arabic{equation}}
\renewcommand{\thetable}{S.\Roman{table}}
 
\setcounter{equation}{0}
\setcounter{figure}{0}
\setcounter{table}{0}
\setcounter{page}{1}
\setcounter{section}{0}

\begin{center}
  \textbf{The topology of SmB$_6$ determined by dynamical mean field theory}\\[.2cm]

{P. Thunstr\"om$^{1,2}$ and  K. Held$^1$}\\[.1cm]

\affiliation{$^1$Institute of Solid State Physics, TU Wien, 1040
Vienna, Austria}
\affiliation{$^2$Department of Physics and Astronomy, Materials Theory, Uppsala University, 751 20 Uppsala, Sweden}

(Dated: \today)\\[1cm]
\end{center}

\begin{center}
\begin{minipage}{.8\textwidth}
In this supplemental material, we first show that the pole expansion Hamiltonian is the proper description of the topological properties in many-body systems and that it is equivalent (not equivalent) to the so-called topological Hamiltonian for a local (non-local) self-energy. Second, we discuss computational details of our density functional theory plus dynamical mean field theory (DFT+DMFT) calculations, using an adapted exact diagonalization as an impurity solver. Third, additional results of the DFT+DMFT electronic structure are presented. Fourth, we demonstrate the robustness of the topological surface states, unless time-reversal symmetry is broken by a magnetic field.
\end{minipage}
\end{center}

\twocolumngrid

\section{Pole expanded Hamiltonian}
\subsection{Local self-energy}
In the following we will show that for a local (e.g. DMFT) self-energy the  pole extended (PE) Hamiltonian \cite{savrasov2006many} and the topological Hamilton \cite{PhysRevX.2.031008} are topologically equivalent. The proof employs some  DMFT concepts explicitly, which can also be used whenever we have a local self-energy (we can just define an impurity model from that local self-energy and the non-interacting Green's function).
In the end of this Section, we will then proof that they are in general not equivalent if non-local self-energies are permitted.

In DMFT each correlated lattice site is mapped to a single impurity Anderson model and the resulting self-energy is mapped back to the lattice. 
The impurity model consists formally of a set of local correlated orbitals, {\em e.g.} the Sm 4f orbitals ($\mathcal{F}$), and a set of non-interacting bath orbitals ($\mathcal{B}$). 
The one-particle term $H$ of the impurity Hamiltonian can hence be partitioned as
\begin{equation}
\label{eqn:hamlocal}
H = H_{FF} + H_{FB} + H_{BF} + H_{BB},
\end{equation}
where the subscripts denote the orthogonal projections upon $\mathcal{F}$ and $\mathcal{B}$, {\em i.e.} $H_{FB} \equiv F H B$ where $F$ and $B$ are the corresponding projection operators. Together, the full projection $P=F+B$ spans the entire system at hand, {\em i.e.} $H_{PP} = H$.

The Lehmann representation of the interacting {\em local} Green's function shows that it can be written as a (large) sum of simple poles
\begin{equation}
G_{mn}(\nu) = \sum_l U^{\phantom{\dagger}}_{ml} \big[\nu - E_l \big]^{-1} U^\dagger_{ln},\label{eqn:glehmann}
\end{equation}
where $E_l$ and $U^{\phantom{\dagger}}_{ml}U^\dagger_{ln}$ give the pole positions and their weights, respectively, and $mn$ runs over all orbitals in the system ($\mathcal{P}$). The poles $E_l$ originate from energy differences between two many-body states, and the (rectangular) matrix $U$ fulfills $\sum_l U^{\phantom{\dagger}}_{ml}U^\dagger_{ln} = \delta_{mn}$ because of the normalization of the Green's function.

The latter property allows us to decompose the matrix $U$ as $U = PW$, where $W$ contains the eigenvectors of a large auxiliary Hamiltonian $\tilde{H}W = W E$ and $P$ is the projection onto the (physical) system as before, with the eigenvalues $E_l$ forming the diagonal matrix $E$. 
Substituting $U = PW$ into Eq.~(\ref{eqn:glehmann}) yields the following expression for the local Green's function
\begin{equation}
G(\nu) = PW \big[\nu - E\big]^{-1} W^\dagger P = P \big[\nu - \tilde{H}\big]^{-1} P.\label{eqn:glehmann2}
\end{equation}
Eq.~(\ref{eqn:glehmann2}) shows that the interacting Green's function takes the form of a projection of the non-interacting Green's function of $\tilde{H}$ onto the physical orbitals. 

The inverse of $G(\nu)$ is required to obtain the self-energy $\Sigma(\nu)$ from the Dyson equation,
\begin{equation}
\Sigma(\nu) = \nu P - H - G^{-1}(\nu) = \nu P - H_{PP} - \big( P \big[\nu - \tilde{H}\big]^{-1} P \big)^{-1},\label{eqn:dyson}
\end{equation}
where $H = H_{PP}$ is the local non-interacting Hamiltonian from Eq.~(\ref{eqn:hamlocal}). To evaluate this expression we must first partition $\tilde{H}$ in a similar way as the Hamiltonian in Eq.~(\ref{eqn:hamlocal}), but this time with respect to the orbitals in $\mathcal{P}$ and a set of auxiliary orbitals ($\mathcal{A}$),
\begin{equation}
\tilde{H} = \tilde{H}_{AA} + \tilde{H}_{AP} + \tilde{H}_{PA} + \tilde{H}_{PP}.
\end{equation}  
Substituting the operator identity (``downfolding'')
\begin{equation}
P \Big[\nu - \tilde{H}\Big]^{-1}\!\!\!\!\! P = \Big[\nu P - \tilde{H}_{PP} - \tilde{H}_{PA} (\nu - \tilde{H}_{AA} )^{-1} \tilde{H}_{AP} \Big]^{-1}\label{eqn:fold}
\end{equation}
into Eq.~(\ref{eqn:dyson}) yields 
\begin{equation}
\Sigma(\nu) = \tilde{H}_{PP} - H_{PP} + \tilde{H}_{PA} (\nu - \tilde{H}_{AA} )^{-1} \tilde{H}_{AP}.\label{eqn:sigmadownfold}
\end{equation}
Hence, the dynamical part of the self-energy, $\tilde{H}_{PA} (\nu - \tilde{H}_{AA} )^{-1} \tilde{H}_{AP}$, corresponds to a hybridization function to the auxiliary orbitals.
The projection in Eq.~(\ref{eqn:sigmadownfold}) is onto both the impurity orbitals and the bath $P = F + B$.  However, the uncorrelated bath orbitals do not have a self-energy, {i.e.} $B\Sigma(\nu) = \Sigma(\nu)B = 0$, which can be shown {e.g.} from the equation of motion or by integrating out the non-interacting bath orbitals. This implies that $\tilde{H}_{PP} = H_{PP} + \Sigma_{FF}(\infty)$ and $\tilde{H}_{BA} = \tilde{H}_{AB} = 0$, which in turn yields
\begin{equation}
\Sigma(\nu) = \Sigma_{FF}(\infty) + \tilde{H}_{FA} (\nu - \tilde{H}_{AA} )^{-1} \tilde{H}_{AF}.\label{eqn:sigmapolesm}
\end{equation}
Eq.~(\ref{eqn:sigmapolesm}) is equivalent to Eq.~(1) in the main text.

In DMFT, the lattice self-energy is replaced with periodically repeated copies of the local self-energy $\Sigma(\nu)$. The bulk lattice Green's function $G_{k}(\nu)$ is hence given by
\begin{align}
G_{k}(\nu) & = \Big[ \nu - H^{k} - \Sigma(\nu) \Big]^{-1}\nonumber\\
& = \Big[ \nu - H^{k} - \Sigma_{FF}(\infty) - \tilde{H}_{FA} (\nu - \tilde{H}_{AA} )^{-1} \tilde{H}_{AF} \Big]^{-1}\label{eqn:gk}
\end{align}
where $k$ belongs to the first Brillouin zone; and the projection operator $F$ corresponds to the Wannier projection of the lattice basis onto the local correlated orbitals. We may now use Eq.~(\ref{eqn:fold}) again, but this time in the other direction (``upfolding''), to extend the Hilbert space with the auxiliary orbitals in $\mathcal{A}$,
\begin{equation}
G_{k}(\nu) = L \Big[ \nu - H^{k} + H_{FF} - \tilde{H}_{(F+A)(F+A)} \Big]^{-1} \! \!\! \! L,\label{eqn:glattice}
\end{equation}
where $L$ projects upon the physical orbitals of the lattice, just as $P$ projected on the physical orbitals (impurity + bath) in Eq.~(\ref{eqn:glehmann2}) in the context of the impurity model. The pole extended (PE) lattice Hamiltonian is hence given by
\begin{equation}
H^{PE}_k \equiv H^{k} + \tilde{H}_{(F+A)(F+A)} - H_{FF}.
\end{equation}

While our derivation assumed a local, i.e., $\mathbf{k}$-independent self-energy,
Eq.~(\ref{eqn:glehmann}) and  Eq.~(\ref{eqn:fold}) can also be formulated for each $\mathbf{k}$-vector. The important difference is that this makes the auxiliary Hamiltonian $\mathbf{k}$-dependent, $\tilde{H}^{k}$, but with this modification Eq.~(\ref{eqn:glattice}) still holds. Hence, both, for a $\mathbf{k}$-independent and a $\mathbf{k}$-dependent self-energy, the physical Green's function, and thus the spectral function, is completely determined by the eigenvalues of $H^{PE}_k$ and the projection onto the correlated orbitals. 

If the non-interacting PE Hamiltonian $H^{PE}_k$ is topologically non-trivial then there will be topological surface states. The physical weights of these surface states will depend on the hybridization between the physical and the auxiliary orbitals at the surface. 
Only a vanishing hybridization can potentially confine a topological surface state of $H^{PE}_k$ to the auxiliary sector, and thus prevent it from appearing in the physical spectrum of $G_{k}(\nu)$.
However, the topological surface states must be dispersive to cross the bulk band gap of $H^{PE}_k$, which implies that if the self-energy is $\mathbf{k}$-independent the topological surface states have to carry a finite physical weight.\\

\subsection{Detaching the auxiliary orbitals}
Two insulating Hamiltonians are topologically equivalent if they can be continuous transformed into each other without closing the band gap during the transformation. Our goal in this section is to show that this type of interpolation can indeed be found between the PE Hamiltonian $H^{PE}_k$ and the topological Hamiltonian \cite{PhysRevX.2.031008} $H_k^T \equiv H^{k} + \Sigma^{k}(0)$ which we combine with the purely auxiliary Hamiltonian $\tilde{H}^{k}_{AA}$ giving $H_k^{TA} \equiv H_k^T + \tilde{H}^{k}_{AA}$, under the condition that both $H^{PE}_k$ and $H_k^{TA}$ have band gaps. This shows that $H^{PE}_k$ and $H_k^T$ are topologically equivalent only as long as $\tilde{H}^{k}_{AA}$ is topologically trivial. 

\begin{figure}[t]
\includegraphics[height=8.cm,angle=90]{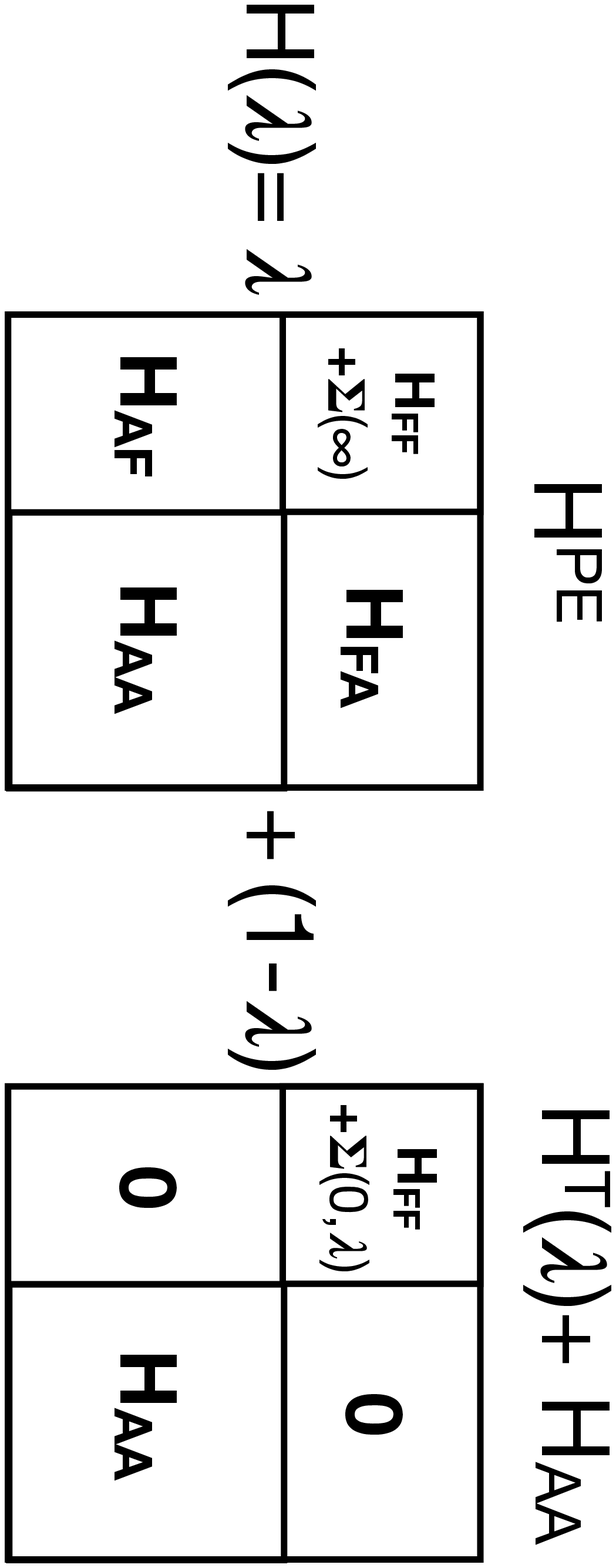}
\caption{Schematics of the interpolation between the topological Hamiltonian at $\lambda=0$ and the PE Hamiltonian at $\lambda=1$ in Eq.~(\ref{eqn:hinter}), which consists of the  correlated ($\mathcal{F}$) physical orbitals and the auxiliary ($\mathcal{A}$) orbitals. Here $\Sigma(0,\lambda) = \Sigma^k_{FF}(\infty)  - (1+\lambda)\tilde{H}^k_{FA} [\tilde{H}^k_{AA}]^{-1} \tilde{H}^k_{AF}$ and $H^T(\lambda) = H^{k} + \Sigma(0,\lambda)$.\label{fig:hinterpol}}
\end{figure}

\begin{figure}
\includegraphics[height=8.cm,angle=90]{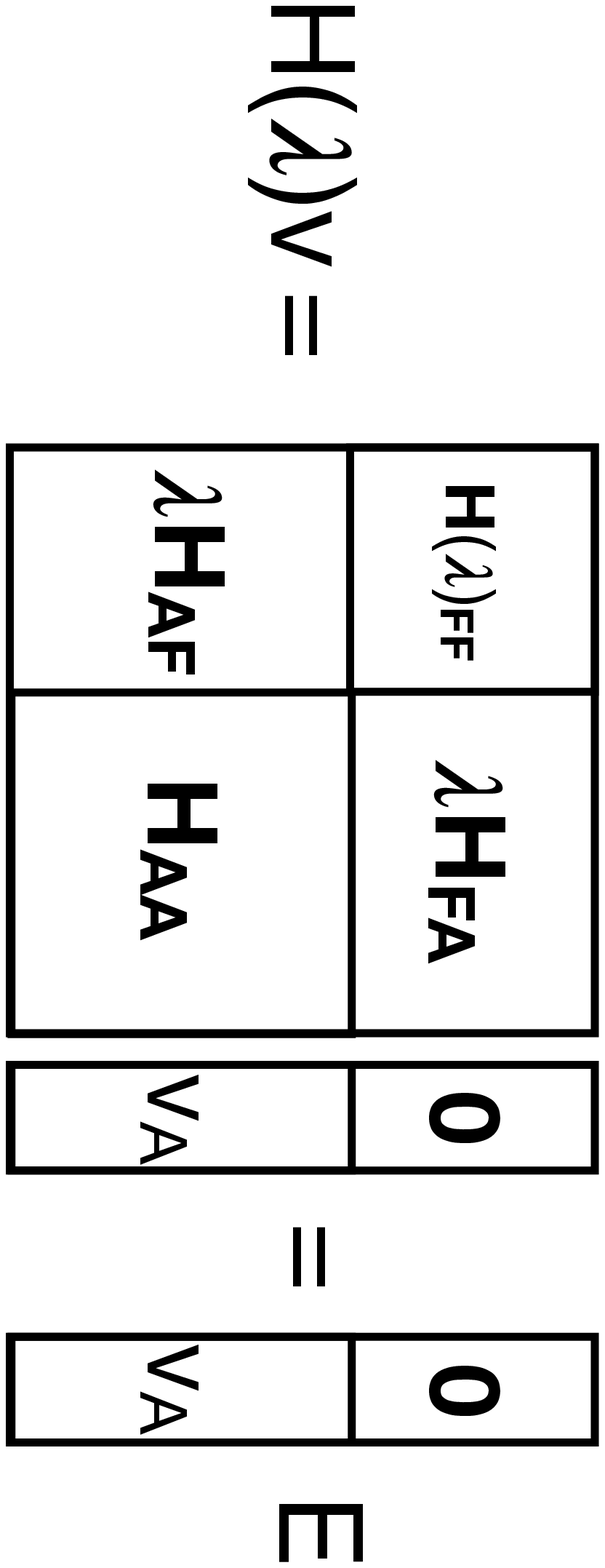}
\caption{For a purely auxiliary eigenvector $v$ the entrance in the physical block is zero and the auxiliary entry is here denoted as $v_A$. Because of this structure $v_A$ is also eigenvector of $H_{AA}$ with eigenvalue $E$ and $H_{FA}\, v_A = 0$ if $\lambda > 0$. Consequently $v$ is also eigenvector of $H_k^{PE} = H(\lambda = 1)$.\label{fig:hargument}}
\end{figure}

The continuous interpolation can be defined as follows (for a visualization, see Fig.~\ref{fig:hinterpol})
\begin{widetext}
\begin{equation}
H^k(\lambda) \equiv H^{k}  + \lambda \Big(\Sigma^k_{FF}(\infty) + \tilde{H}^k_{(F+A)(F+A)} - \tilde{H}^k_{FF}\Big) + (1-\lambda)\Big(\Sigma^k_{FF}(\infty)  - (1+\lambda)\tilde{H}^k_{FA} [\tilde{H}^k_{AA}]^{-1} \tilde{H}^k_{AF} + \tilde{H}^k_{AA}\Big) 
,\label{eqn:hinter}
\end{equation}
so that $H^k(\lambda=0) = H^{T}_k$ and $H^k(\lambda=1) = H^{PE}_k$. The corresponding physical lattice Green's function becomes
\begin{equation}
G^\lambda_{k}(\nu)  =  L \Big[ \nu - H^{k}(\lambda) \Big]^{-1} L =  \Big[ \nu - H^{k} - \Sigma^k_{FF}(\infty) - \tilde{H}^k_{FA} \Big(\lambda^2[\nu A - \tilde{H}^k_{AA} ]^{-1} - (1-\lambda^2)[\tilde{H}^k_{AA}]^{-1} \Big) \tilde{H}^k_{AF} \Big]^{-1}\!\!\!\!\!.\label{eqn:ginter}
\end{equation}
\end{widetext}
A key point is that interpolation is chosen in such a way that the physical Green's function evaluated at the Fermi energy, $G^\lambda_{k}(0)$, is independent of $\lambda$, 
\begin{equation}
G^\lambda_{k}(0) = \Big[ -H^{k} -\Sigma^k_{FF}(\infty) + \tilde{H}^k_{FA} [\tilde{H}^k_{AA}]^{-1} \tilde{H}^k_{AF} \Big]^{-1}\!\!\!\!\!.
\end{equation}
The end-point systems, i.e., the PE ($\lambda = 1$) and the topological Hamiltonian with detached auxiliary orbitals ($\lambda = 0$), have band gaps. Since further $G^\lambda_{k}(0)$ is independent of $\lambda$, no band with a finite physical weight can cross the Fermi energy during the interpolation, as this would affect $G^\lambda_{k}(0)$. The only remaining possibility to close the band gap is if a purely auxiliary band crosses the Fermi energy, {i.e.} that there exists an eigenvector $v$ such that $v = A v$ and $H^k(\lambda) v = 0$ for some $k$ and $1 > \lambda > 0$. However, since $v$ has no physical ($\mathcal{F}$) component, $H^k(\lambda)_{FA} v = \lambda\tilde{H}_{FA} v = 0$, see Fig.~\ref{fig:hargument}. Furthermore, the auxiliary part of $H^k(\lambda)$ is identical to that of the PE Hamiltonian, $H^k(\lambda)_{AA} = (H^{PE}_k)_{AA}$, which together with $\tilde{H}_{FA} v = 0$ gives $H^{PE}_k v = 0$. That is, the purely auxiliary vector $v$ must also be an eigenvector of $H^{PE}$ with a zero eigenvalue. This eigenvector can not exist since the PE Hamiltonian was assumed to be insulating. This implies that no band can cross the Fermi energy during the continuous interpolation in Eq.~(\ref{eqn:hinter}). 

The PE Hamiltonian $H^{PE}_k$ and the combined topological auxiliary Hamiltonian $H_k^{TA}$ are hence topologically equivalent. The topological Hamiltonian $H_k^{T}$ is by itself only topologically equivalent to $H^{PE}_k$ if the detached auxiliary Hamiltonian $\tilde{H}^k_{AA}$ is topologically trivial. $\tilde{H}^k_{AA}$ is automatically trivial if the self-energy is local ($k$-independent), but for a non-local self-energy it becomes necessary to consider how the auxiliary orbitals contribute to the elementary band representations \cite{nature23268}, as shown by two concrete examples in the next section.

\subsection{Non-local self-energy}
Let us illustrate the topological role of the non-local self-energy by constructing two simple examples for which PE and topological Hamiltonian are not topologically equivalent. From the definition of the topological Hamiltonian $H_k^T \equiv H^{k} + \Sigma^{k}(0)$ it is clear that if $\Sigma^{k}(0)$ is much smaller than the band gap of $H^{k}$ it cannot change the topology of $H_k^T$. On the other hand, the auxiliary orbitals in the PE Hamiltonian may change its topology even with an infinitesimal coupling strength. Let us therefore start with a system governed by the topologically non-trivial Bernevig-Hughes-Zhang (BHZ) Hamiltonian\cite{Bernevig2006quantum}
\begin{widetext}
\begin{equation}
H_{BHZ}^k = \left(
\begin{array}{cccc}
 1-\cos (k_x)-\cos (k_y) & 0.5 (\sin (k_x)-i \sin (k_y)) & 0 & 0 \\
 0.5 (\sin (k_x)+i \sin (k_y)) & -1+\cos (k_x)+\cos (k_y) & 0 & 0 \\
 0 & 0 & 1-\cos (k_x)-\cos (k_y) & -0.5 (\sin (k_x)+i \sin (k_y)) \\
 0 & 0 & -0.5 (\sin (k_x)-i \sin (k_y)) & -1+\cos (k_x)+\cos (k_y) \\
\end{array}
\right),
\end{equation}
\end{widetext}
and define a non-local self-energy as $\Sigma^k(\nu) = |V|^2 [\nu + H_{BHZ}^k]^{-1}$, where $V$ is the coupling strength. Such a self-energy can be obtained if we have a non-local and frequency-dependent interaction. The corresponding PE and topological Hamiltonians can be easily identified as
\begin{align}
H^{PE}_k & = \left(
\begin{array}{cc}
 H_{BHZ}^k & V\mathbf{1} \\
 V^*\mathbf{1} & -H_{BHZ}^k \\
\end{array}
\right),\label{eqn:ex1a}\\
H^{T}_k & = H_{BHZ}^k + |V|^2 [H_{BHZ}^k]^{-1},\label{eqn:ex1b}
\end{align}
where $\mathbf{1}$ is the identity matrix, and the second row and column in the matrix in Eq.~\ref{eqn:ex1a} corresponds to the auxiliary orbitals ({c.f.} Fig.~\ref{fig:hinterpol}). $H^{T}_k$ is topologically non-trivial for all $V$, since $[H_{BHZ}^k]^{-1}$ has the same eigenvectors as $H_{BHZ}^k$ and the sign of the eigenvalues are identical. On the other hand, the reversed sign of $H_{BHZ}^k$ in the auxiliary sector of $H^{PE}_k$ makes the PE system topologically trivial for all $V$ \footnote{The physical and auxiliary orbitals form two separated non-trivial subspaces when $V=0$. However, since $V=0$ is not considered a protected symmetry, this separation is not topological and the full system is still trivial.}. This proves that the PE and topological Hamiltonian are topologically not equivalent.

\begin{figure}
\noindent\begin{tabular}{p{4.2cm} p{0.5mm} p{4.2cm} }
\centering $H^{PE}_k$ from Eq.~({\ref{eqn:ex1a}}) & & \centering $H^{T}_k$ from Eq.~({\ref{eqn:ex1b}}) 
\end{tabular}
\includegraphics[width=4.2cm]{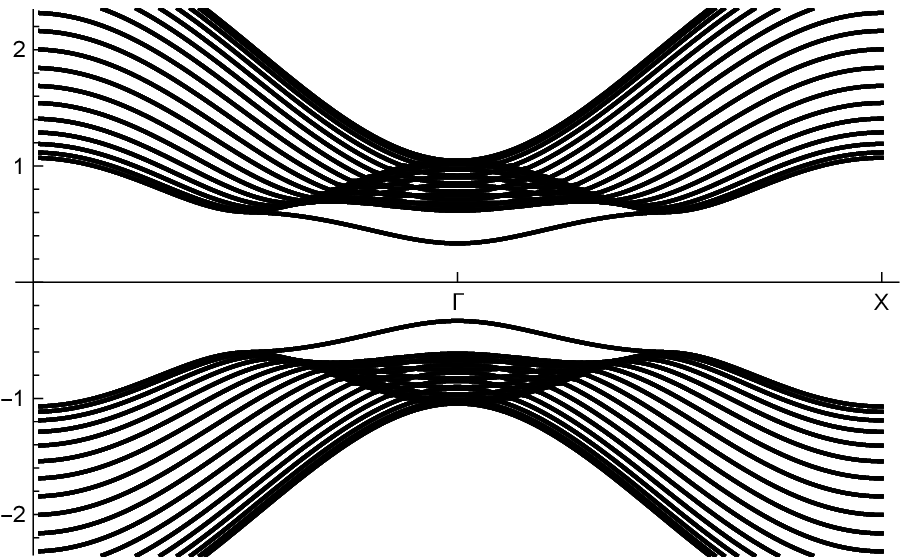}\hspace{0.5mm}\includegraphics[width=4.2cm]{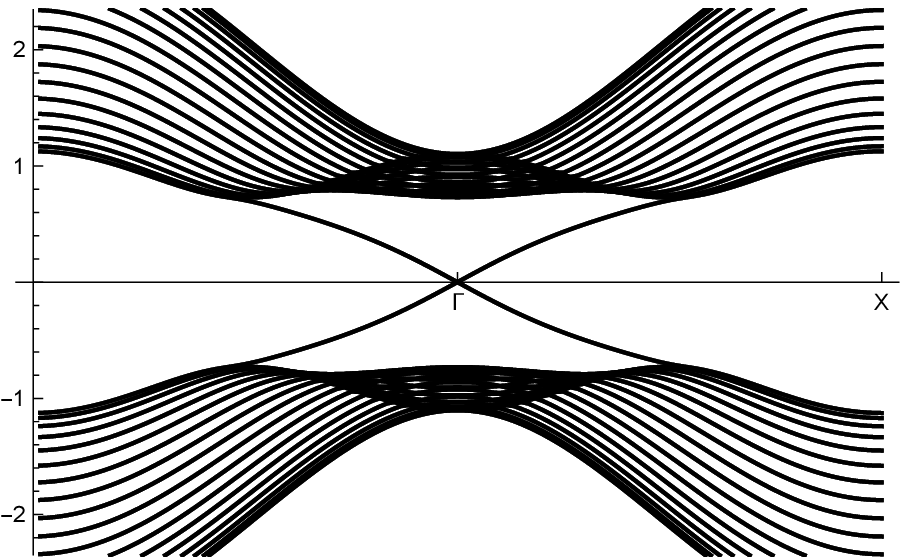}\\
\vspace{2mm}\noindent\begin{tabular}{p{4.2cm} p{0.5mm} p{4.2cm} }
\centering $H^{PE}_k$ from Eq.~({\ref{eqn:ex2a}}) & & \centering $H^{T}_k$ from Eq.~({\ref{eqn:ex2b}}) 
\end{tabular}
\includegraphics[width=4.2cm]{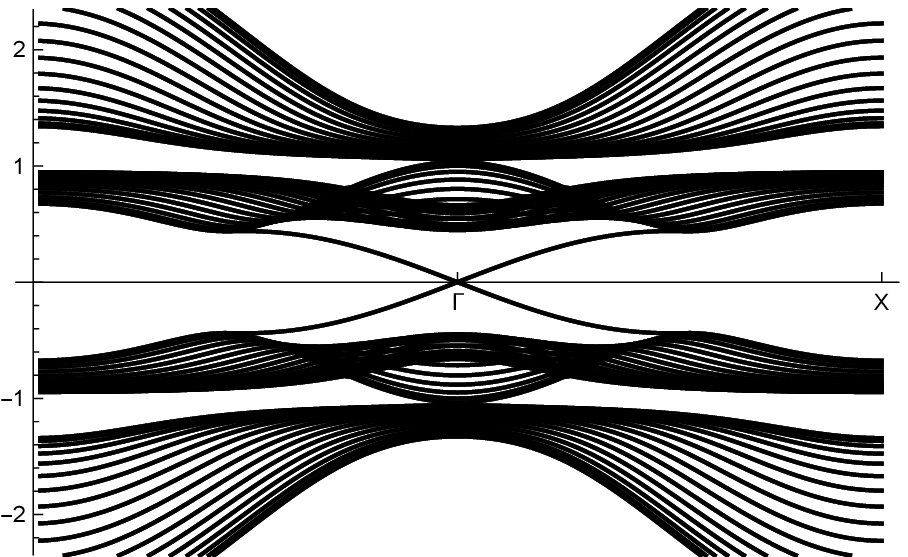}\hspace{0.5mm}\includegraphics[width=4.2cm]{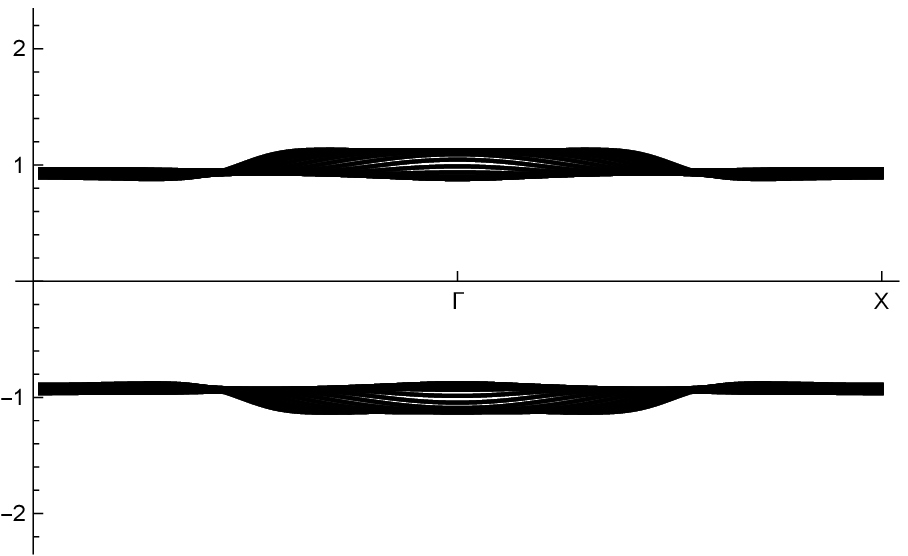}
\caption{Slab calculations with non-local self-energies. The $k$-resolved eigenvalues of the PE Hamiltonian (left) and the topological Hamiltonian (right) for an 18 layer slab, with $V = 1/3$. The upper panel shows the Hamiltonians in Eq.~({\ref{eqn:ex1a}}) and Eq.~({\ref{eqn:ex1b}}), and the lower panel shows the Hamiltonians in Eq.~({\ref{eqn:ex1a}}) and Eq.~({\ref{eqn:ex1b}}).
In both examples with a non-local self-energy, the  topological Hamiltonian ($H^T_k$) gives the wrong topology.
\label{fig:bhzslab}}
\end{figure}

This is confirmed by slab calculations, see Fig.~\ref{fig:bhzslab}, using $H^{PE}$ from Eq.~({\ref{eqn:ex1a}}) and $H^{T}$ from Eq.~({\ref{eqn:ex1b}}). These reveal the presence of metallic surface states for $H^{T}$ while the spectrum is gapped for $H^{PE}$, as expected.

Another example is given by a system with the same non-local self-energy but with the constant Hamiltonian
\begin{equation}
H^{loc} = \left(
\begin{array}{cccc}
 -1& 0 & 0 & 0 \\
 0 & 1 & 0 & 0 \\
 0 & 0 &-1 & 0 \\
 0 & 0 & 0 & 1 \\
\end{array}
\right).
\end{equation}
The corresponding PE and topological Hamiltonians become
\begin{align}
H^{PE}_k & = \left(
\begin{array}{cc}
 H_{loc} & V\mathbf{1} \\
 V^*\mathbf{1} & -H_{BHZ}^k \\
\end{array}
\right),\label{eqn:ex2a}\\
H^{T}_k & = H_{loc} + |V|^2 [H_{BHZ}^k]^{-1}\label{eqn:ex2b}.
\end{align}
The PE Hamiltonian of Eq.~(\ref{eqn:ex2a}) is non-trivial if $|V|^2 < 1$, while it becomes trivial when $|V|^2 > 3$. The topological Hamiltonian has just the opposite classification.

These two examples clearly demonstrates the topological role of the auxiliary orbitals and the poles of the self-energy. Even when it is possible to detach the auxiliary orbitals from the physical system, their contribution to the topology cannot be neglected. 

One may make an analogy to a system with spin-orbit coupling, and let the spin-up and the spin-down channel take the role of the physical and the auxiliary orbitals, respectively. Even if it would be possible to smoothly switch off the spin-orbit coupling, and hence detach the two spin channels, one can still not determine the topology of the original spin-full system (PE Hamiltonian) from the spin-up channel alone (topological Hamiltonian). Just as the topological invariants must trace over both spin channels, they must also at least formally trace over the auxiliary orbitals.

To sum up, we have proven that the PE and topological Hamiltonian are topologically equivalent for a local ({e.g.} DMFT) self-energy but may be topologically distinct for a non-local self-energy. Only the PE Hamiltonian properly reflects the topology of the system, whereas for a non-local self-energy the topological Hamiltonian may yield the wrong topology.

\section{Computational details}
All the SmB$_6$ DFT+DMFT calculations were performed using the experimental cubic crystal structure (space group: 221, prototype: CaB$_6$) with the lattice parameter $a = 4.13$\AA. The DFT exchange-correlation functional was set to the local density approximation\cite{perdew1992accurate}, and the Brillouin zone was sampled through a conventional Monkhorst-Pack mesh of 16 x 16 x 16 k-points in the bulk and 16 x 16 x 1 k-points in the slab calculations. The local Coulomb interaction was parameterized in terms of the Slater parameters $F^0$, $F^2$, $F^4$, and $F^6$. The Hubbard $U$ parameter $F^0$ is heavily screened by the valence electrons and was set to $8.0$ eV. The less screened Slater parameters $F^2$, $F^4$, and $F^6$ were instead calculated through Slater integrals at the beginning of each DFT iteration, and then scaled by 0.95, 0.97, and 1.00, respectively~\cite{hitoshi2012high}.The final self-consistent values are given in Table \ref{tab:slater}.

The fully charge self-consistent DFT+DMFT calculation were performed using the full-potential linear muffin-tin orbital (LMTO) code {\scriptsize RSPt} \cite{Wills00} and the DMFT implementation presented in Refs.~\onlinecite{Grechnev07, granas2012charge, thunstrom2012electronic}. The lattice Hamiltonian as well as the DMFT lattice Green's function are calculated in the full LMTO basis. The projection operators upon the localized Sm 4f orbitals are obtained from a $\mathbf{k}$-dependent L{\"o}wdin orthogonalization of the LMTO SM 4f orbitals, as detailed in the Supplemental Material of Ref.~\cite{thunstrom2012electronic}.
\begin{table}
\caption{Screened Sm 4f Slater integrals (F$^0$ -- F$^6$) and the double counting potential ($\mu_{DC}$) of the Sm atoms in the bulk and the Sm terminated (Sm) and B$_6$ terminated (B$_6$) supercells in Fig.~\ref{fig:slab-band}. The labels start from the middle Sm atom (Sm1) and go toward the surface (Sm2 -- Sm4). The Sm atoms have an intermediate valence except the outer most Sm atom which is Sm$^{2+}$ (Sm$^{3+}$) in the slab with Sm (B$_6$) termination. The fully localized limit double counting potential ($\mu_{DC}^{FLL}$) is given as a reference. All values are given in eV. \label{tab:slater}}
\begin{tabular}{c c @{~~~} c c c c @{~~~} c c}
\hline
Label & Termination & F$_0$ & F$_2$ & F$_4$ & F$_6$ & $\mu_{DC}$ & $\mu_{DC}^{FLL}$\\
\hline
Sm  & Bulk   & 8.0 & 11.43 & 7.49 & 5.54 & 35.32 & 37.75\\
\hline
Sm1 & Sm    & 8.0 & 11.39 & 7.47 & 5.52 & 35.55 & 37.82\\
Sm2 & Sm    & 8.0 & 11.40 & 7.47 & 5.52 & 35.55 & 37.91\\
Sm3 & Sm    & 8.0 & 11.39 & 7.47 & 5.52 & 35.70 & 37.85\\
Sm4 & Sm    & 8.0 & 10.83 & 7.07 & 5.22 & 39.62 & 41.75\\
\hline
Sm1 & B$_6$ & 8.0 & 11.43 & 7.50 & 5.54 & 35.25 & 37.72\\
Sm2 & B$_6$ & 8.0 & 11.43 & 7.50 & 5.54 & 35.26 & 37.67\\
Sm3 & B$_6$ & 8.0 & 11.42 & 7.49 & 5.54 & 35.29 & 37.75\\
Sm4 & B$_6$ & 8.0 & 11.86 & 7.80 & 5.77 & 32.33 & 34.37\\
\hline
\end{tabular}
\end{table}

\subsection{Exact diagonalization}
The Sm 4f orbitals are strongly contracted compared to the Sm 5d and 6s orbitals, which makes the Sm 4f orbitals hybridize very weakly with the orbitals on neighboring atoms. The hybridization is completely captured by the local hybridization function $\Delta(\nu)$ \cite{georges1996dynamical,kotliar2006electronic,held2007electronic} that describes how the electrons propagate in the material once they leave the Sm 4f orbitals of a given atom. To put the weak Sm 4f hybridization in perspective, the Ni 3d hybridization function in NiO is about 20 times larger than the Sm 4f hybridization function in SmB$_6$ for comparable computational setups. The electronic structure of SmB$_6$ is therefore already well-described on the eV energy scale by completely neglecting the hybridization \cite{thunstrom2009multiplet}. However, this approach is not accurate enough to describe the important meV energy scale close to the Fermi energy. It is particularly important that the hybridization function at the Fermi energy $\Delta(0)$ is accurately captured, as well as the Sm 4f occupation, to ensure that the Fermi energy remains in the hybridization band gap during the DFT+DMFT self-consistency cycle. The Exact Diagonalization (ED) impurity solver \cite{caffarel94exact,thunstrom2012electronic,liebsch2011temperature} takes most of the hybridization between the correlated Sm 4f orbitals and the rest of the material into account by including a limited number of effective bath orbitals in the impurity problem. However, the total number of bath orbitals that can be included in the impurity problem is severely limited by the growth of the many-body Hilbert space. In order to still get an accurate representation of $\Delta(0)$ and the Sm 4f occupation we need to go beyond the standard hybridization fitting scheme as well as modifying the double counting correction. The details thereof are described in the next two subsections.

\subsection{Bath discretization}
In the self-consistent DMFT scheme the lattice Green's function in Eq.~(\ref{eqn:gk}) is projected onto the local Sm 4f orbitals ($\mathcal{F}$) and integrated over the Brillouin zone to yield the local Green's function $G_{FF}(\nu)$. The hybridization function is then extracted from the inverse of $G_{FF}(\nu)$,
\begin{equation}
\Delta(\nu) = \nu\mathbf{1}_{FF} - H_{FF} - \Sigma_{FF}(\nu) - [G_{FF}(\nu)]^{-1},
\end{equation}
where $H_{FF}$ is the local impurity Hamiltonian in Eq.~(\ref{eqn:hamlocal}) which includes the spin-orbit coupling, the double-counting, and all crystal field terms. $\Sigma_{FF}(\nu)$ is the local self-energy given by Eq.~(\ref{eqn:sigmapolesm}). The ED method proceeds by fitting a few bath state parameters $H_{FB}$ and $H_{BB}$ to $\Delta(\nu)$ via the model function
\begin{equation}
\Delta^{ED}(\nu) = H_{FB}[\nu\mathbf{1}_{BB} - H_{BB}]^{-1}H_{BF}.
\end{equation}
We fit the hybridization function on the Matsubara axis with 6 bath states per Sm orbital using a conjugate gradient scheme which also takes the off-diagonal terms in $\Delta(\nu)$ into account. The high-energy bath states ($\mathcal{B}'$), {i.e.} the eigenstates of $H_{BB}$ with energies $E_{b'}$ relatively far from the Fermi energy ($|E_{b'}| \gg w_{b'} \equiv \sqrt{H_{b'F}H_{Fb'}}$), give only a perturbative contribution to the low energy physics due to the large energy cost of exciting these bath states. Their effect can be estimated by introducing the scaling $H_{B'B'} \rightarrow \lambda H_{B'B'}$ and $H_{FB'} \rightarrow \sqrt{\lambda} H_{FB'}$ in $\Delta^{ED}(\nu)$, which keeps $\Delta^{ED}(0)$ invariant, and let the scaling parameter $\lambda \rightarrow \infty$. The scaling shows that the high energy bath states can be replaced to zeroth order by a static contribution
\begin{align}
\Delta^{ED}(\nu) \approx & H_{FB''}[\nu\mathbf{1}_{B''B''} - H_{B''B''}]^{-1}H^\dagger_{B''F} \nonumber\\
& + H_{FB'}[-H_{B'B'}]^{-1}H^\dagger_{B'F},
\end{align}
where $B''$ is the low-energy complement to $B'$, i.e.~$B = B'' + B'$. In our calculations we put the high-energy cut-off at $|E_{b'}| > 10 w_{b'}$. The remaining low-energy bath states were ordered according to their weight $w_{b}$, and included in $\mathcal{B}''$ to the extent allowed by the computational resources, with a minimum of 4 bath states in total. The converged one-particle term of the ED Hamiltonian $H^{ED}_0 = H_{FF} + H_{FB} + H_{BF} + H_{BB}$ for the bulk calculation is presented in Fig.~\ref{fig:hed}.
\begin{figure*}
$\left(
\begin{array}{ @{\!} c @{\!\!\!} c @{\!\!\!} c @{\!\!\!} c @{\!\!\!} c @{\!\!\!} c @{\!\!\!} c @{\!\!\!} c @{\!\!\!} c @{\!\!\!} c @{\!\!\!} c @{\!\!\!} c @{\!\!\!} c @{\!\!\!} c @{\!\!\!\!\!} c @{\!\!\!\!\!} c @{\!\!\!\!\!} c @{\!\!\!\!\!} c @{\!}}
 \!\!-35.8918\!\! & 0. & 0. & 0. & -0.0041 & 0. & 0. & 0. & 0. & 0. & -0.0014 & 0. & 0. & 0. & 0. & 0. & 0. & 0. \\
 0. & \!\!-35.9871\!\! & 0. & 0. & 0. & 0. & 0. & 0.1905 & 0. & 0. & 0. & 0.015 & 0. & 0. & -0.1932 & 0. & 0.2735 & 0. \\
 0. & 0. & \!\!-36.0769\!\! & 0. & 0. & 0. & 0.0136 & 0. & 0.2748 & 0. & 0. & 0. & 0.015 & 0. & 0. & 0.2735 & 0. & -0.0082 \\
 0. & 0. & 0. & \!\!-36.1422\!\! & 0. & 0. & 0. & 0. & 0. & 0.2844 & 0. & 0. & 0. & -0.0014 & 0. & 0. & 0. & 0. \\
 -0.0041 & 0. & 0. & 0. & \!\!-36.2184\!\! & 0. & 0. & 0. & 0. & 0. & 0.2844 & 0. & 0. & 0. & 0. & 0. & 0. & 0. \\
 0. & 0. & 0. & 0. & 0. & \!\!-36.3163\!\! & 0. & -0.0109 & 0. & 0. & 0. & 0.2748 & 0. & 0. & -0.1537 & 0. & -0.264 & 0. \\
 0. & 0. & 0.0136 & 0. & 0. & 0. & \!\!-36.398\!\! & 0. & -0.0109 & 0. & 0. & 0. & 0.1905 & 0. & 0. & -0.2122 & 0. & 0.0068 \\
 0. & 0.1905 & 0. & 0. & 0. & -0.0109 & 0. & \!\!-36.398\!\! & 0. & 0. & 0. & 0.0136 & 0. & 0. & -0.2122 & 0. & 0.0068 & 0. \\
 0. & 0. & 0.2748 & 0. & 0. & 0. & -0.0109 & 0. & \!\!-36.3163\!\! & 0. & 0. & 0. & 0. & 0. & 0. & -0.1537 & 0. & -0.264 \\
 0. & 0. & 0. & 0.2844 & 0. & 0. & 0. & 0. & 0. & \!\!-36.2184\!\! & 0. & 0. & 0. & -0.0041 & 0. & 0. & 0. & 0. \\
 -0.0014 & 0. & 0. & 0. & 0.2844 & 0. & 0. & 0. & 0. & 0. & \!\!-36.1422\!\! & 0. & 0. & 0. & 0. & 0. & 0. & 0. \\
 0. & 0.015 & 0. & 0. & 0. & 0.2748 & 0. & 0.0136 & 0. & 0. & 0. & \!\!-36.0769\!\! & 0. & 0. & 0.2735 & 0. & -0.0082 & 0. \\
 0. & 0. & 0.015 & 0. & 0. & 0. & 0.1905 & 0. & 0. & 0. & 0. & 0. & \!\!-35.9871\!\! & 0. & 0. & -0.1932 & 0. & 0.2735 \\
 0. & 0. & 0. & -0.0014 & 0. & 0. & 0. & 0. & 0. & -0.0041 & 0. & 0. & 0. & \!\!-35.8918\!\! & 0. & 0. & 0. & 0. \\ 
 0. & -0.1932 & 0. & 0. & 0. & -0.1537 & 0. & -0.2122 & 0. & 0. & 0. & 0.2735 & 0. & 0. & -3.3062 & 0. & 0. & 0. \\
 0. & 0. & 0.2735 & 0. & 0. & 0. & -0.2122 & 0. & -0.1537 & 0. & 0. & 0. & -0.1932 & 0. & 0. & -3.3062 & 0. & 0. \\
 0. & 0.2735 & 0. & 0. & 0. & -0.264 & 0. & 0.0068 & 0. & 0. & 0. & -0.0082 & 0. & 0. & 0. & 0. & -1.796 & 0. \\
 0. & 0. & -0.0082 & 0. & 0. & 0. & 0.0068 & 0. & -0.264 & 0. & 0. & 0. & 0.2735 & 0. & 0. & 0. & 0. & -1.796 \\
\end{array}
\right)$
\caption{The one-particle term $H^{ED}_0$, including the double counting correction and the down-folded high-energy bath states, of the self-consistent Sm 4f ED Hamiltonian of SmB$_6$ (bulk). The 14 Sm 4f spin-orbitals are located in the upper-left corner and ordered according to $(l_z,s_z) = (-3,-1/2),(-2,-1/2),\cdots,(3,-1/2),(-3,1/2),\cdots,(3,1/2)$. The remaining 4 spin-orbitals in the lower right corner are the bath states. The energy unit is eV.\label{fig:hed}}
\end{figure*}

\subsection{Double counting correction}
In the DFT+DMFT scheme, the explicit addition of a local Coulomb interaction term to the DFT Hamiltonian introduces a double counting (DC) of the electron-electron interaction. The unknown form of the screening processes in the exchange correlation functional prevents the implementation of an exact double counting correction. In particular, the screening of the local interaction between the Sm 4f electrons, implicitly described within the local density approximation, may be different to the effective (static) screening implied by the renormalized Slater parameters. Due to this ambiguity several different double counting corrections schemes have been suggested over the years, such as the Fully Localized Limit (FLL) \cite{solovyev1994corrected} and Around Mean Field (AMF) \cite{anisimov1991band}. The FLL correction removes the spherically averaged Hartree-Fock contribution of an effective atomic-like system with integer orbital occupations. The AMF correction considers instead an effective itinerant system with uniform (non-integer) orbital occupations. However, the thermal ground state of an intermediate valence compound such as Sm$_6$ falls outside these two scenarios as it contains several thermally occupied almost atomic-like many-body eigenstates having different number of electrons, as shown in Table~\ref{tab:gs}. The occupation of an intermediate valence ground state responds to the DC potential in Fermi-Dirac-like steps. However, the underlying assumptions of both FLL and AMF make their DC correction linear in the occupation. The two different behaviors always lead to a feed-back loop away from intermediate valence towards integer valence. A second less sever issue is that the finite discretization of the bath states in ED causes a small mismatch between the impurity green's function and the local green's function $G_{FF}(\nu)$ projected from the lattice. To minimize these two problems we automatically adjusted the double counting potential at each DMFT iteration to obtain the same number of Sm 4f electrons in the impurity as in the lattice. In the bulk calculation the number of Sm 4f electrons stabilizes at 5.48 at a temperature of 100 K, remarkable close to the experimental value of 5.45\cite{mizumaki2009temperature}. In the slab calculations we chose to enforce the bulk Sm 4f occupation of 5.48 for all the Sm atoms except at the surface, to make the center Sm atom as bulk-like as possible and minimize the hybridization between the surface states located at the top and bottom of the slab. 

\section{Many-body electronic structure}
\subsection{Thermal ground state}
\begin{table}
\caption{The lowest energy many-body eigenstates of the impurity Hamiltonian. The symmetry (Sym.), Sm 4f occupation ($N_f$), energy (E), thermal weight (GS) are tabulated for each eigenstate, as well as their contribution to the Sm 4f total (J), angular (L), and spin (S) moment.\label{tab:gs}}
\begin{tabular}{c c c c c c c c c c c}
\hline
Sym & $N_f$ & E(meV) & GS(\%) & $J$ & $L$ & $S$ & $J_z$ & $L_z$ & $S_z$ \\
\hline
$\Gamma_1^+$ & 6.002 & 0  & 46 & 0.03 & 2.95 & 2.94 &  0.00 &  0.00 &  0.00\\
$\Gamma_8^-$ & 5.030 & 11 & 12 & 2.52 & 4.93 & 2.47 &  1.80 &  3.16 & -1.36\\
$\Gamma_8^-$ & 5.030 & 11 & 12 & 2.52 & 4.93 & 2.47 &  0.47 &  0.75 & -0.28\\
$\Gamma_8^-$ & 5.030 & 11 & 12 & 2.52 & 4.93 & 2.47 & -0.47 & -0.75 &  0.28\\
$\Gamma_8^-$ & 5.030 & 11 & 12 & 2.52 & 4.93 & 2.47 & -1.80 & -3.16 &  1.36\\
$\Gamma_7^-$ & 5.029 & 24 &  3 & 2.53 & 4.93 & 2.47 &  0.77 &  1.07 & -0.30\\
$\Gamma_7^-$ & 5.029 & 24 &  3 & 2.53 & 4.93 & 2.47 & -0.77 & -1.07 &  0.30\\
$\Gamma    $ & 6.002 & 47 &  0 & 1.01 & 2.96 & 2.95 & -1.00 & -0.50 & -0.50\\
$\Gamma    $ & 6.002 & 47 &  0 & 1.01 & 2.96 & 2.95 &  0.00 &  0.00 & -0.00\\
$\Gamma    $ & 6.002 & 47 &  0 & 1.01 & 2.96 & 2.95 &  1.00 &  0.50 &  0.50\\
\hline
\end{tabular}
\end{table}
The lowest energy many-body eigenstates of the self-consistent impurity problem Hamiltonian $H^{ED}$ is given in Table~\ref{tab:gs}. 
The states have almost atomic-like Sm 4f occupations ($N_f$) and total angular momenta ($J$), but the crystal fields and the hybridization with the bath mix the different $J_z$ configurations. The thermal groundstate is an incoherent mixture of the eigenstates according to their Boltzmann weights (GS) $e^{-\beta E}/Tr[e^{-\beta H}]$ at 100 K. At this temperature the largest contributions are given by the four degenerate $\Gamma_8^-$ states (48\%) with $J \approx 5/2$ and $N_f \approx 5$, and the $\Gamma_1^+$ state (46\%) with $J \approx 0$ and $N_f \approx 6$, in agreement with non-resonant inelastic X-ray scattering data \cite{Sundermann2018}. The remaining 6\% belongs to a $\Gamma_7^-$ doublet. 

\subsection{Extended photoemission spectrum}
\begin{figure}
\includegraphics[width=0.45\textwidth]{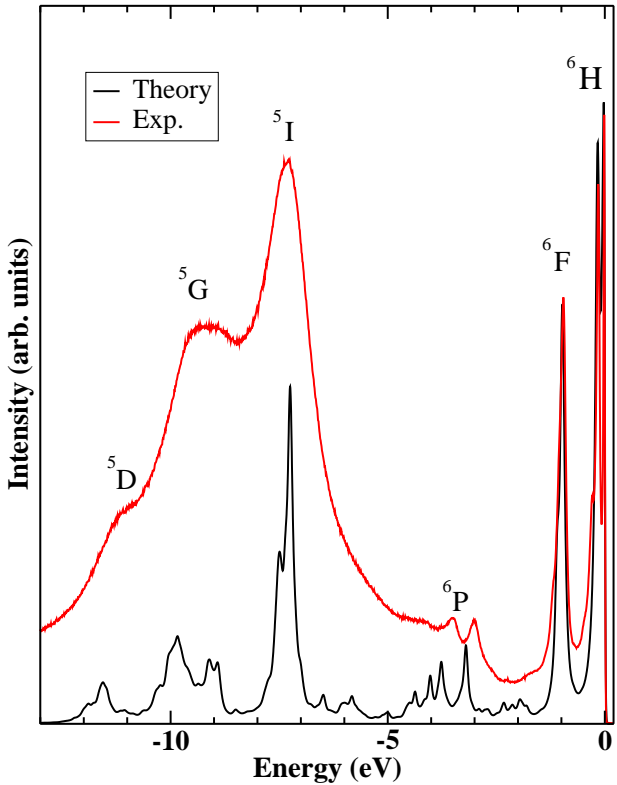}
\caption{The Sm $4f$-projected DFT+DMFT spectral density of SmB$_6$ compared to experimental on-resonance ($\hbar\nu =$ 140 eV) photoemission data\cite{Denlinger2013}. The approximate $J$ and $L$ quantum numbers of the final states are indicated. 
\label{fig:bulk-dos}} 
\end{figure}

In Fig.~\ref{fig:bulk} of the main paper, we have already shown the spectral function in a small energy range from -1.2~eV to 0.1~eV, which resolves the $J$  and $j_z$ multiplets for $L=5$ and $L=3$. In Fig.~\ref{fig:bulk-dos} we show the same $\mathbf k$-integrated spectral function in a larger energy window. Clearly all peaks agree well with the experimentally measured (on-resonance) photoemission data \cite{Denlinger2013}. The relative weight of the $4f^6\rightarrow 4f^5$ transitions are accurately captured. The relative weight of the different $4f^5\rightarrow 4f^4$ transitions are in reasonable agreement with the experimental data, even though we do not include the resonance effect which strongly enhances their total weight. This gives us further assurance that our DFT+DMFT calculation faithfully describe bulk SmB$_6$.

\section{Robustness of the topological surface states}
\subsection{Adding a surface potential}
\begin{figure*}
\noindent\begin{tabular}{p{0.44\textwidth} p{0.44\textwidth}}
\centering {\large\bf Sm 4f sub-surface} & \centering {\large\bf Sm 4f subsub-surface} 
\end{tabular}
\includegraphics[width=0.44\textwidth]{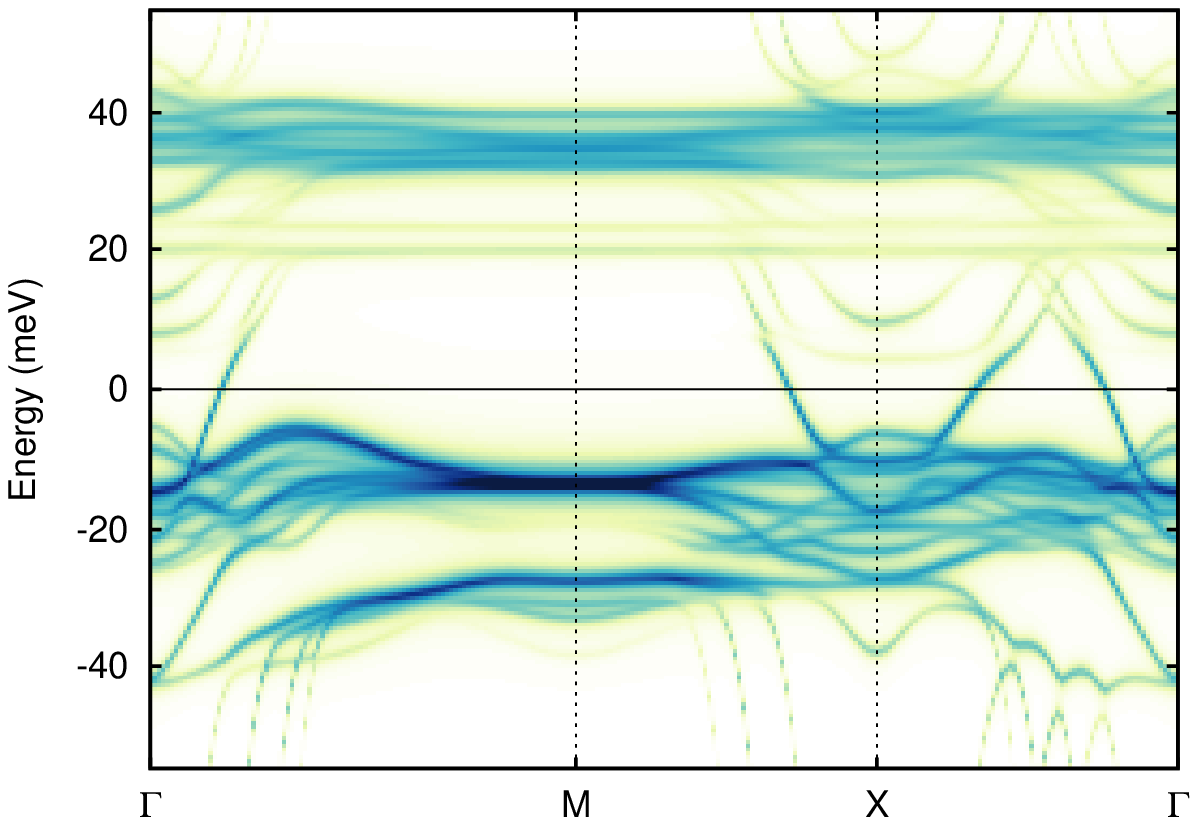}
\includegraphics[width=0.44\textwidth]{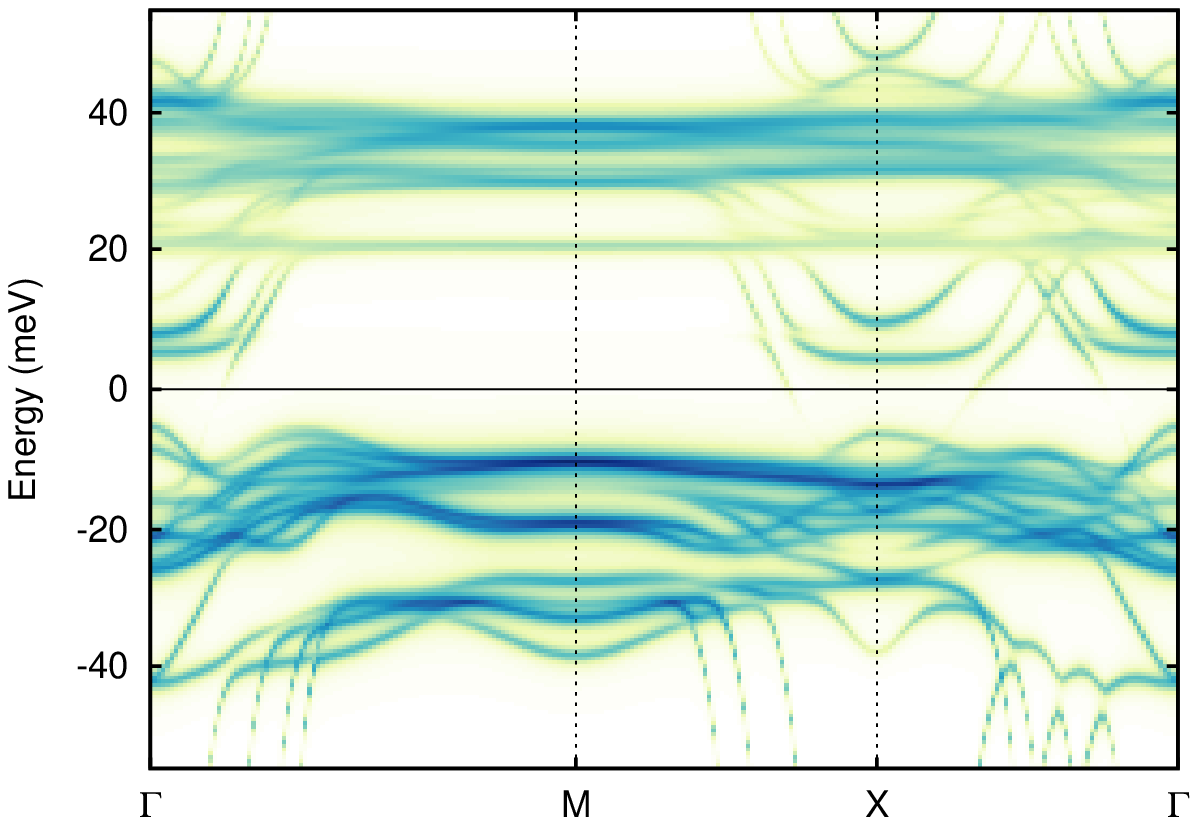}\\
\includegraphics[width=0.44\textwidth]{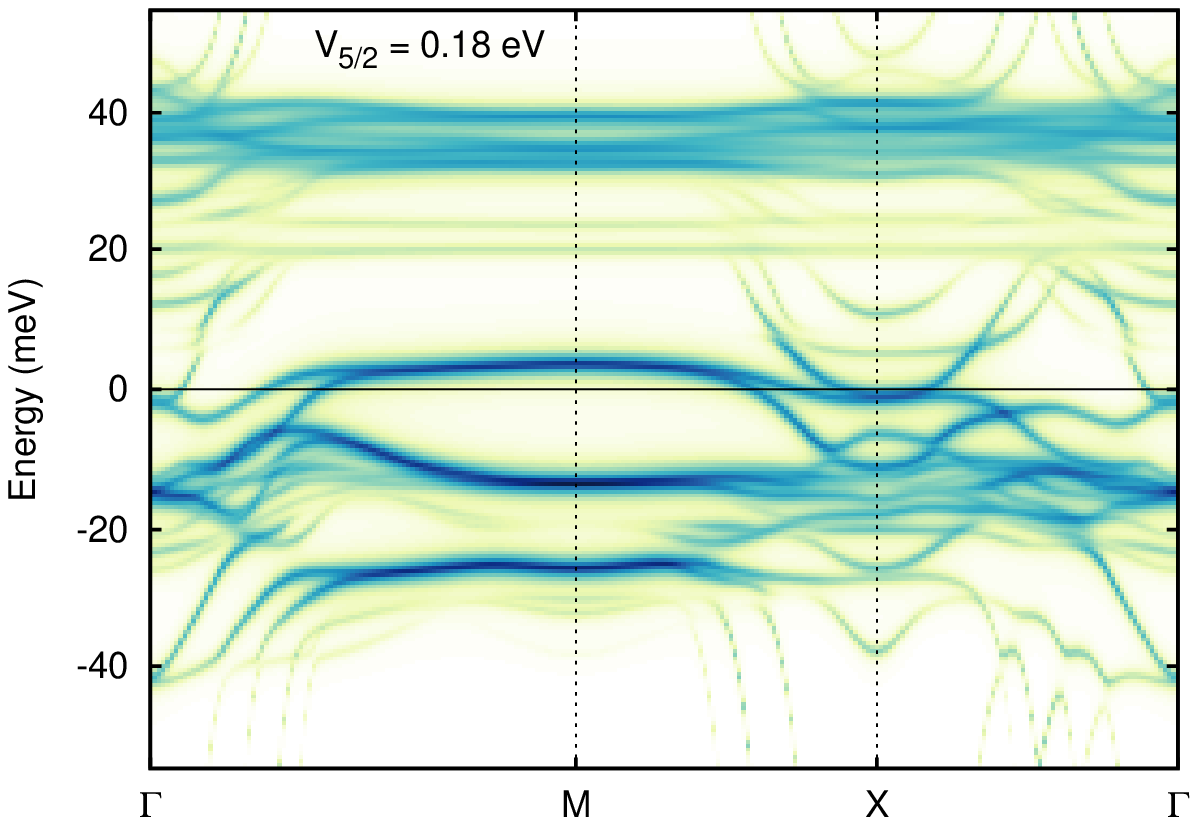}
\includegraphics[width=0.44\textwidth]{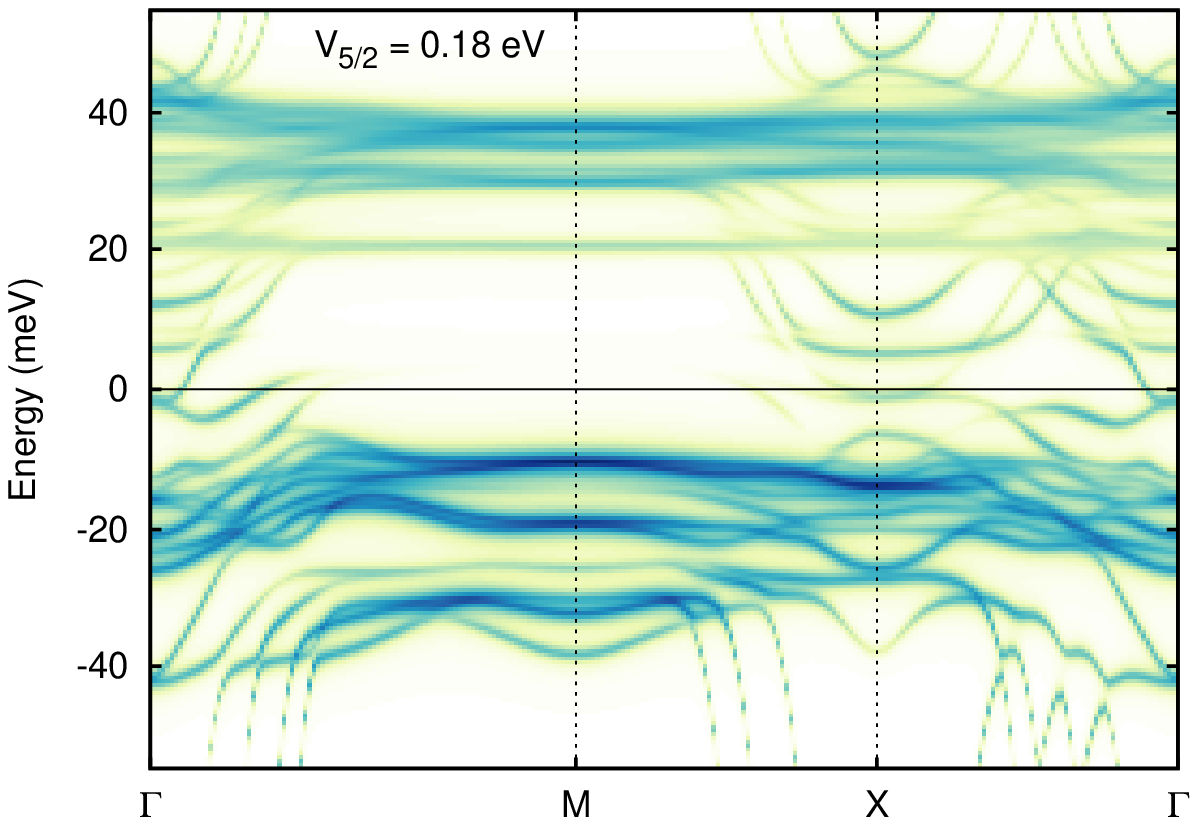}
\includegraphics[width=0.44\textwidth]{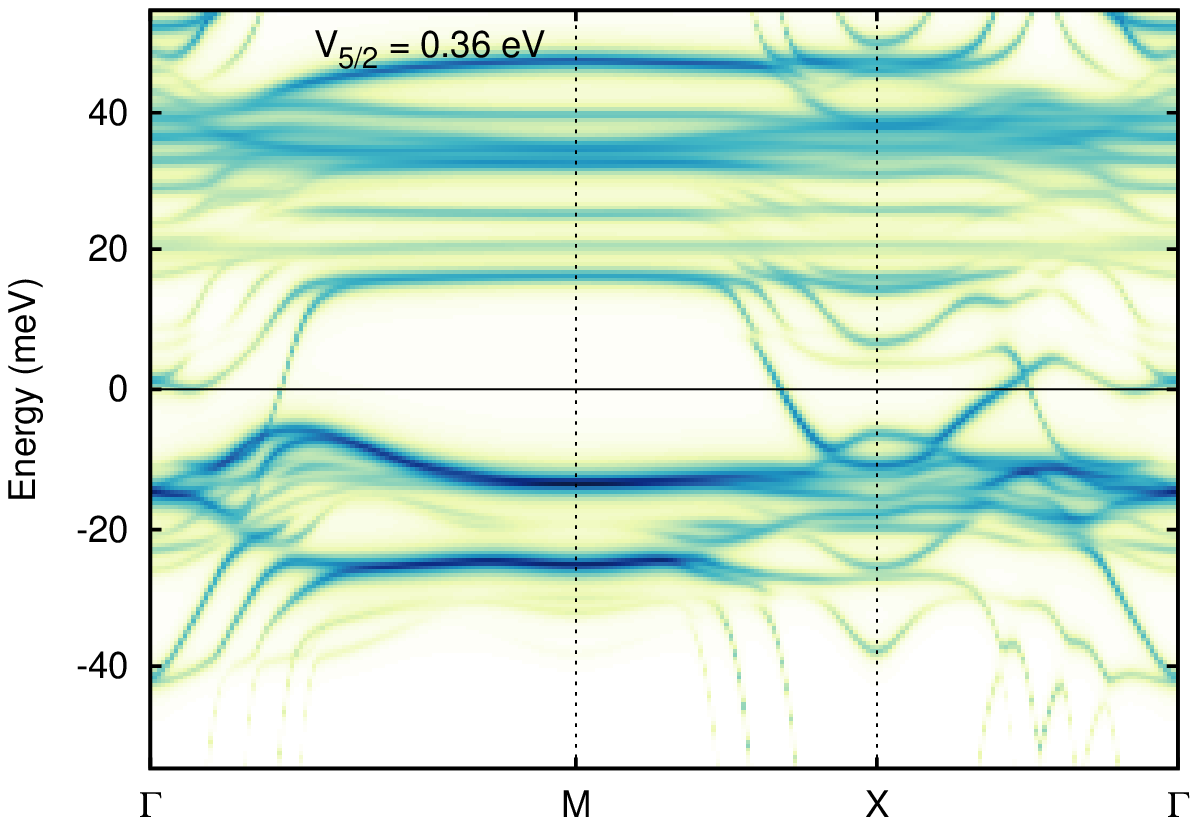}
\includegraphics[width=0.44\textwidth]{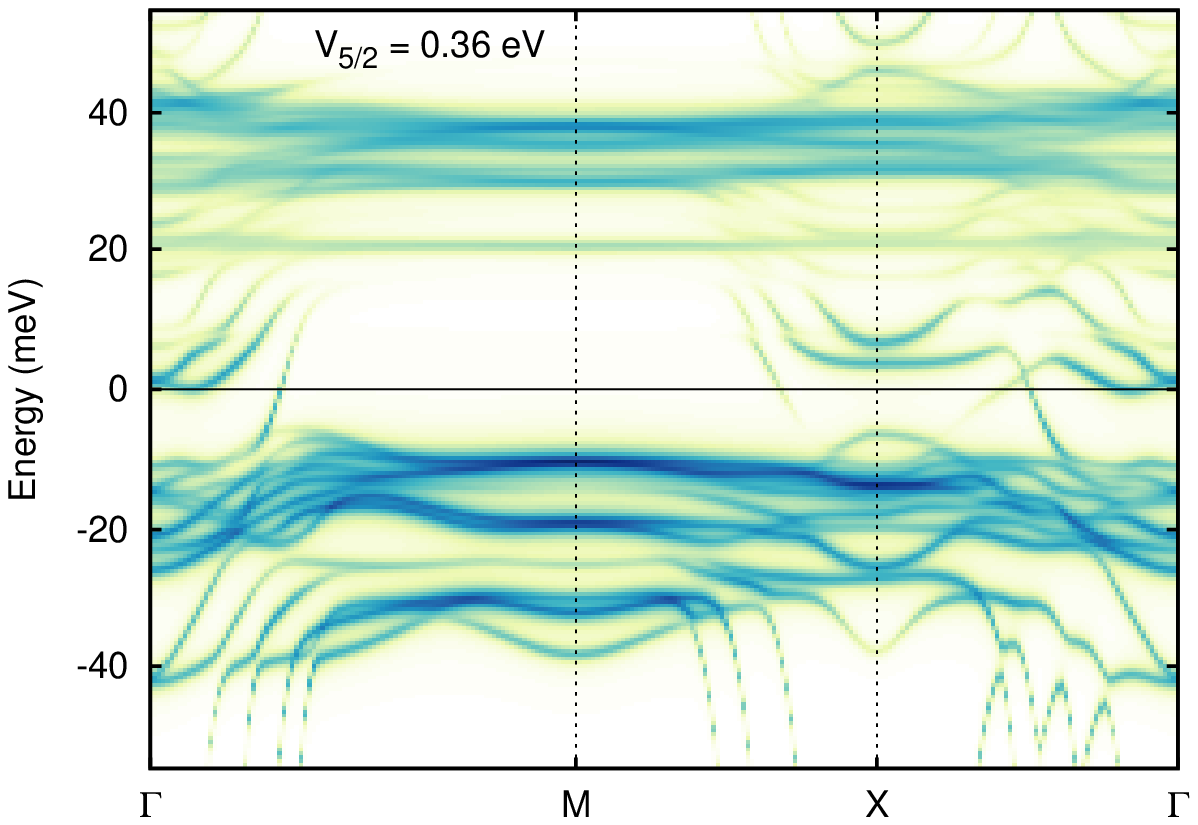}
\includegraphics[width=0.44\textwidth]{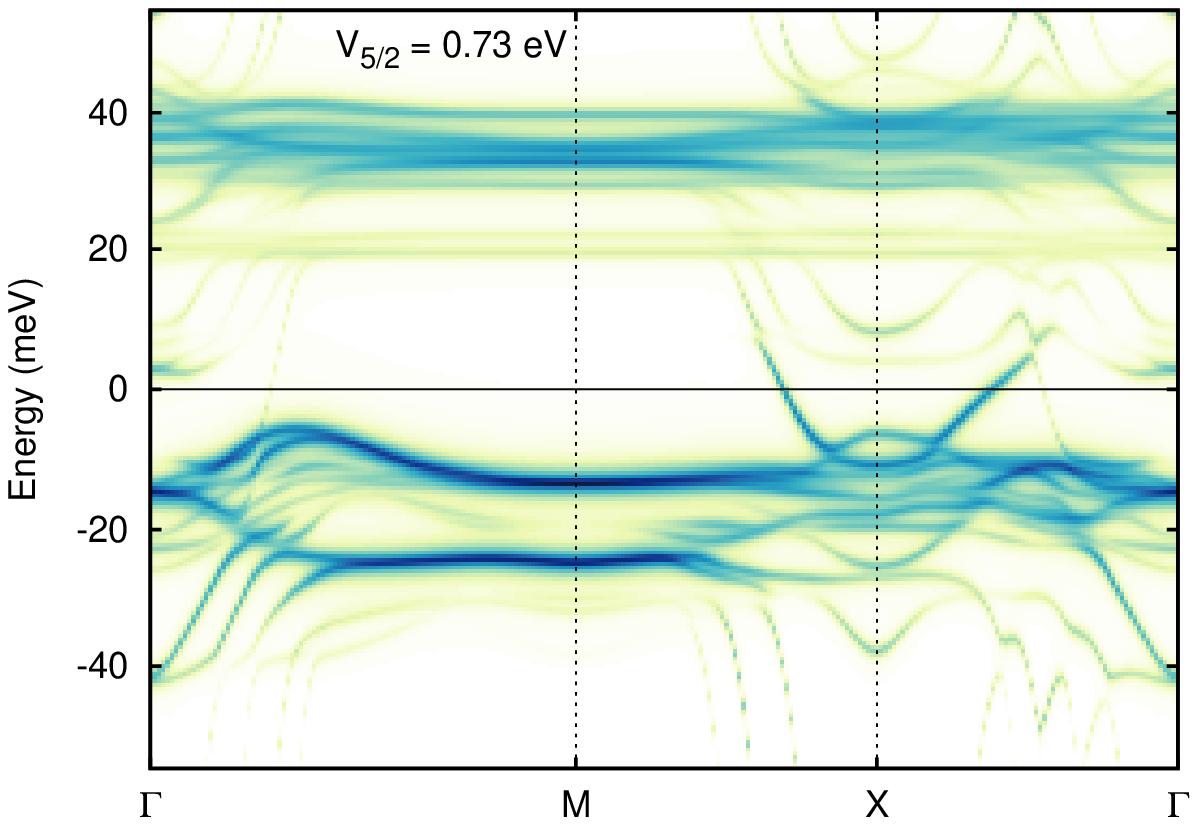}
\includegraphics[width=0.44\textwidth]{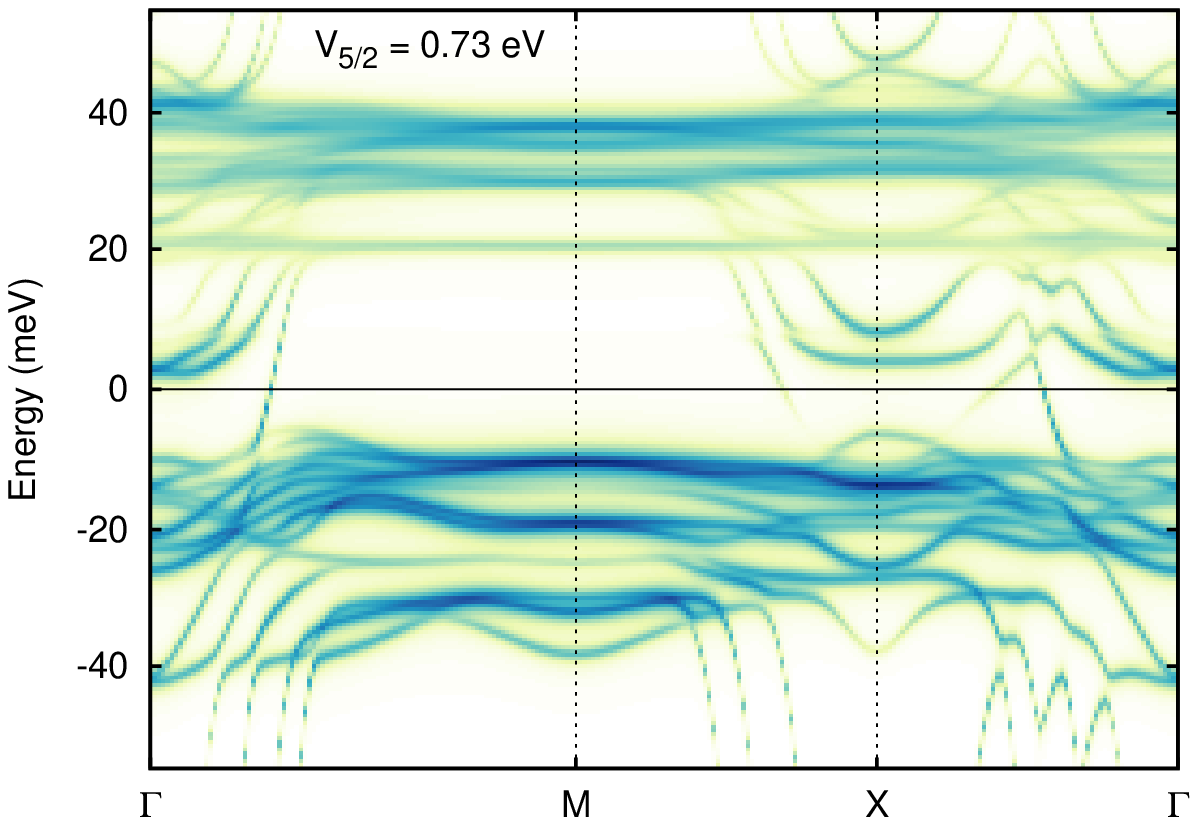}
\caption{
The $\mathbf{k}$-resolved DFT+DMFT spectrum in the presence of a perturbing potential at the sub-surface Sm 4f $J=5/2$, $j_z = \pm5/2$ states of the Sm terminated SmB$_6$ supercell depicted in Fig.~\ref{fig:slab-band}. The left (right) panel shows the projection upon the sub-surface (subsub-surface) Sm $4f$ states. The topological surface states around the $\Gamma$-point shift from the sub-surface to the subsub-surface  as the potential is increased.
\label{fig:slabpotg}}
\end{figure*}

\begin{figure*}
\noindent\begin{tabular}{p{0.45\textwidth} p{0.45\textwidth}}
\centering {\large\bf Sm $4f$ sub-surface} & \centering {\large\bf Sm $4f$ subsub-surface} 
\end{tabular}
\includegraphics[width=0.44\textwidth]{smb6-sub-0.eps}
\includegraphics[width=0.44\textwidth]{smb6-subsub-0.eps}\\
\includegraphics[width=0.44\textwidth]{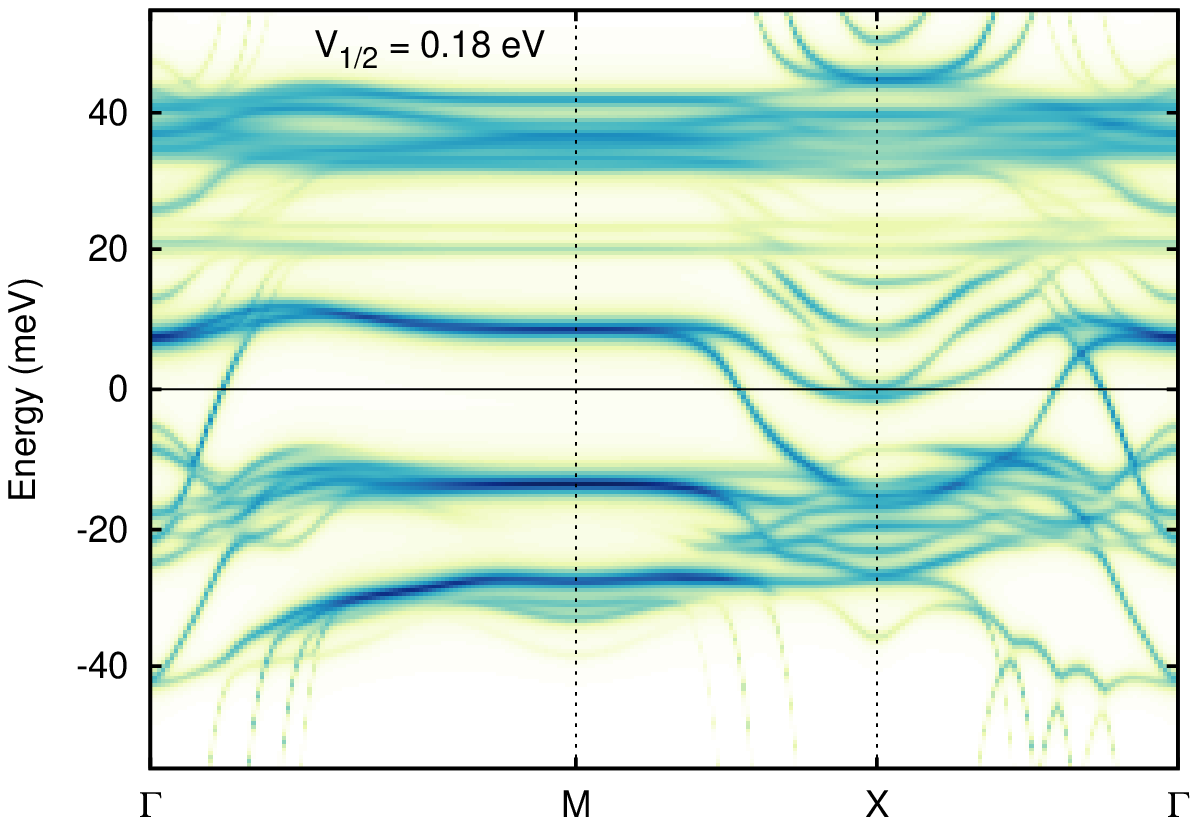}
\includegraphics[width=0.44\textwidth]{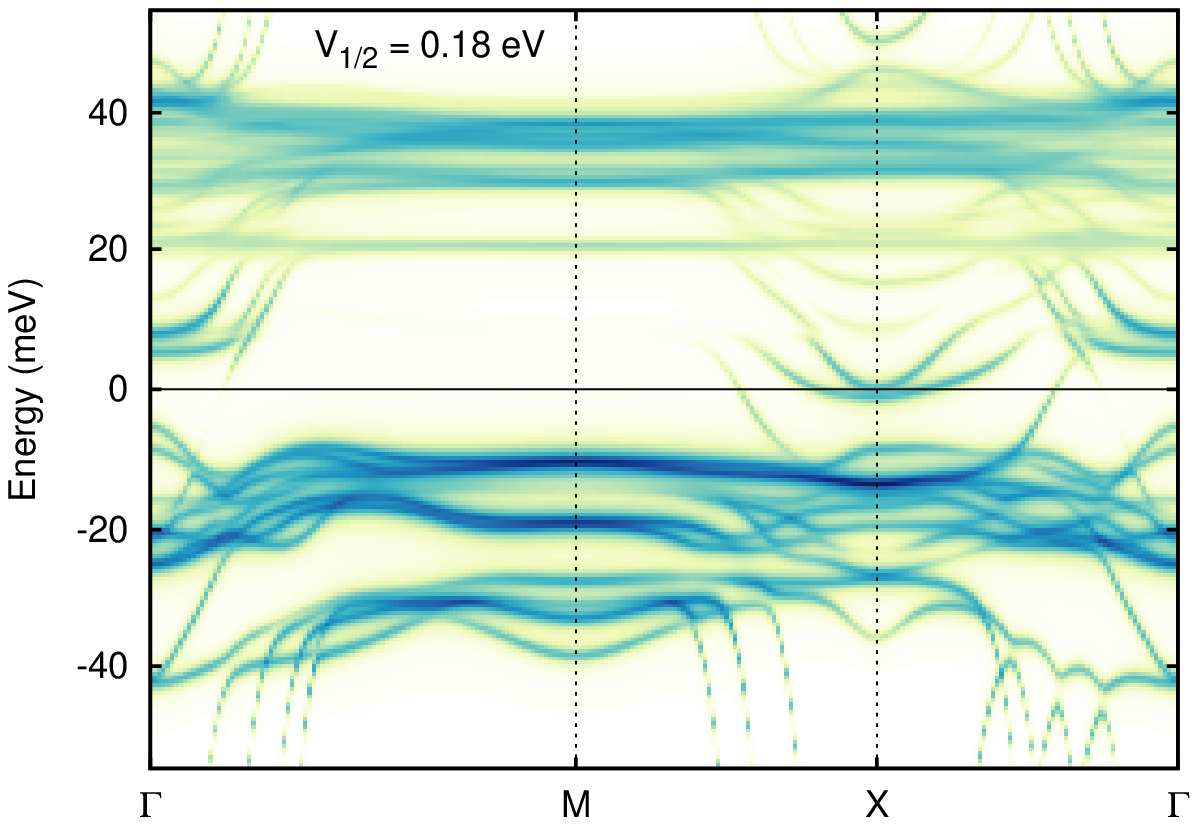}
\includegraphics[width=0.44\textwidth]{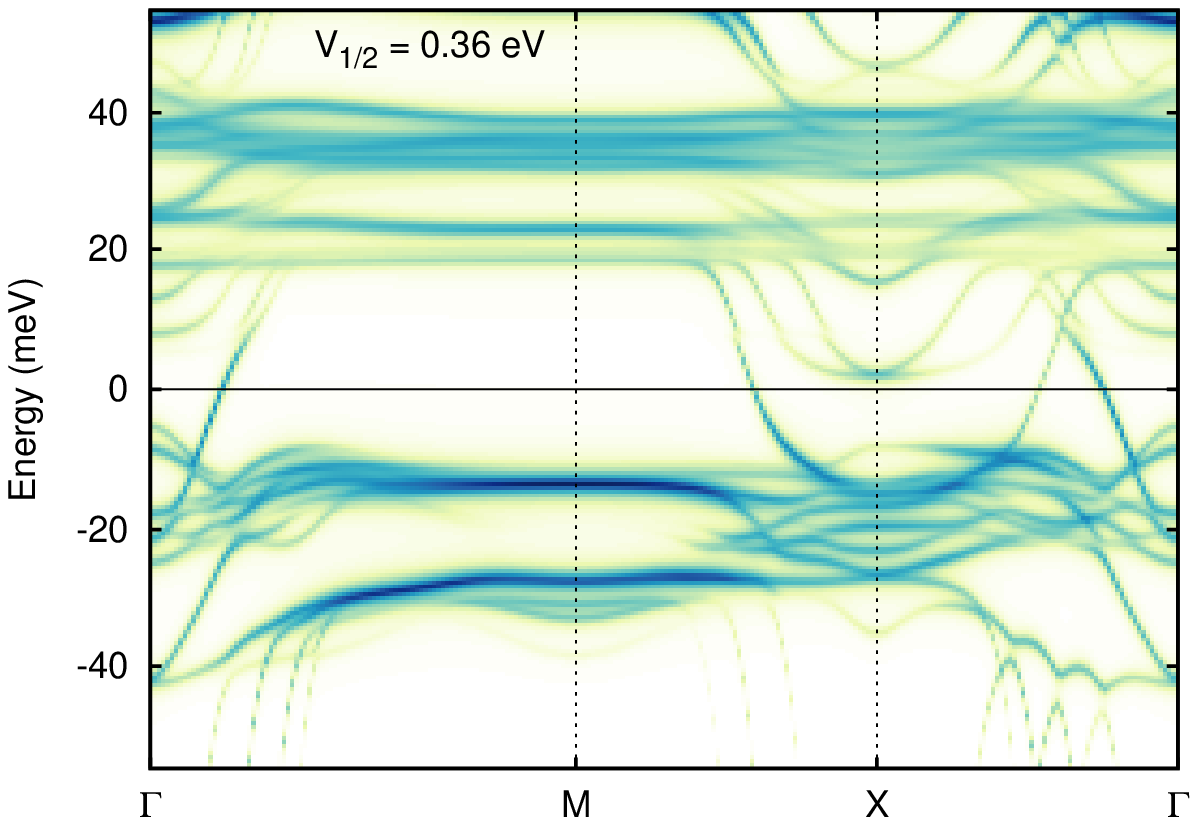}
\includegraphics[width=0.44\textwidth]{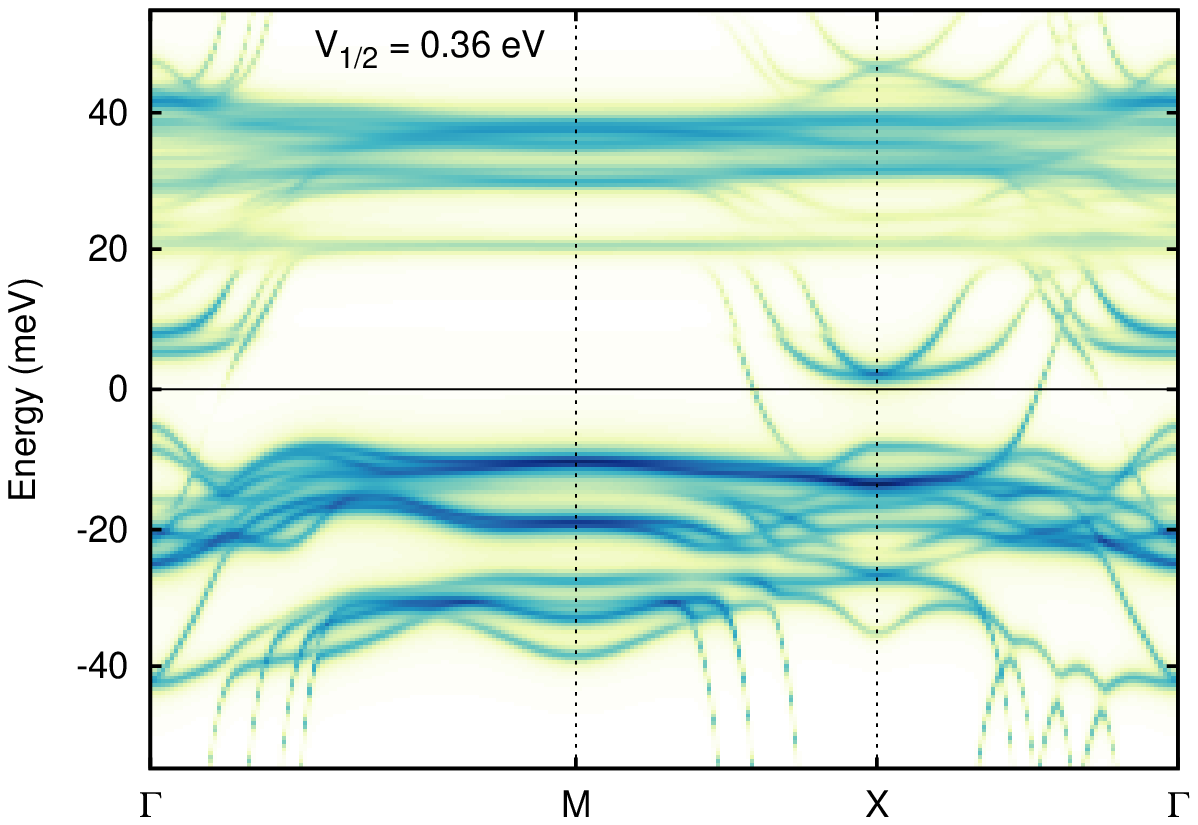}
\includegraphics[width=0.44\textwidth]{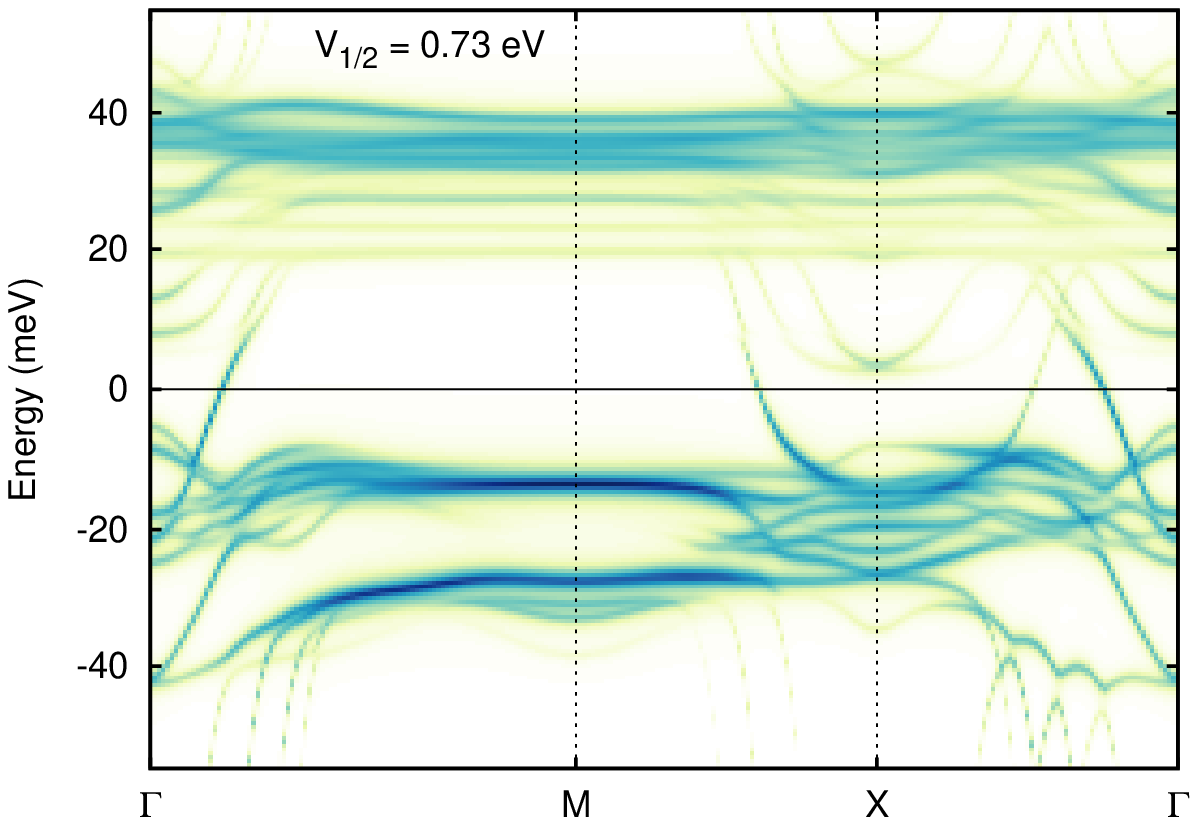}
\includegraphics[width=0.44\textwidth]{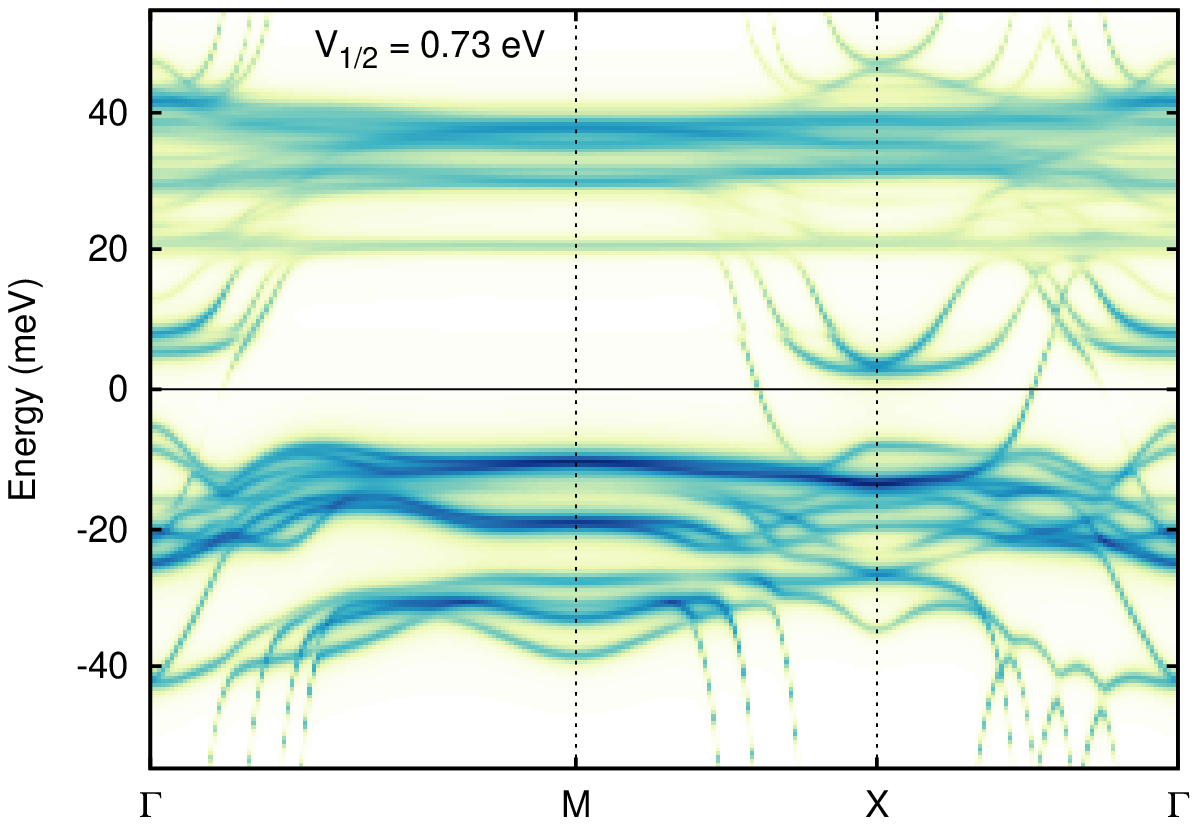}
\caption{
Same as Fig.~\ref{fig:slabpotg}, but now the perturbing potential is applied to the sub-surface Sm 4f $J=5/2$, $j_z = \pm1/2$ states. The topological surface states around the $X$-point shift partially from the sub-surface to the subsub-surface as the potential is increased.
\label{fig:slabpotx}}
\end{figure*}

\begin{figure*}[t!]
\includegraphics[width=0.4\textwidth]{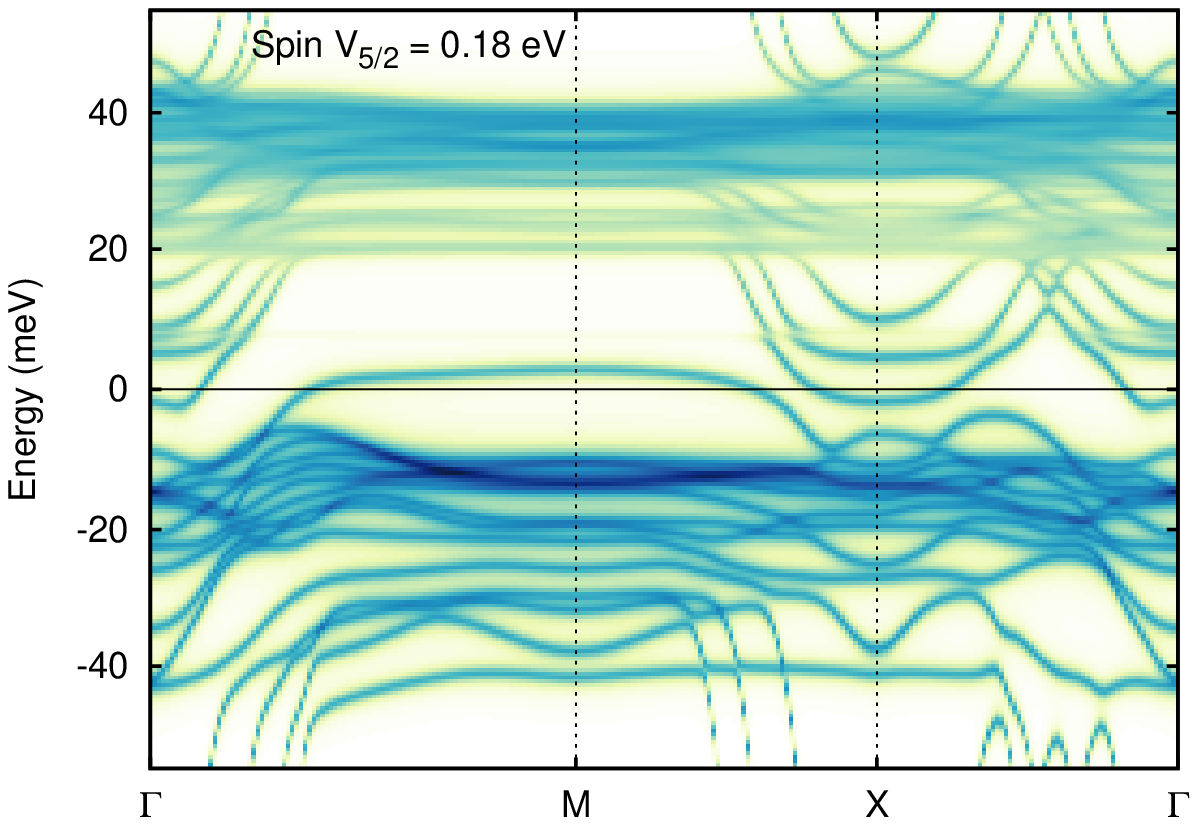}
\includegraphics[width=0.4\textwidth]{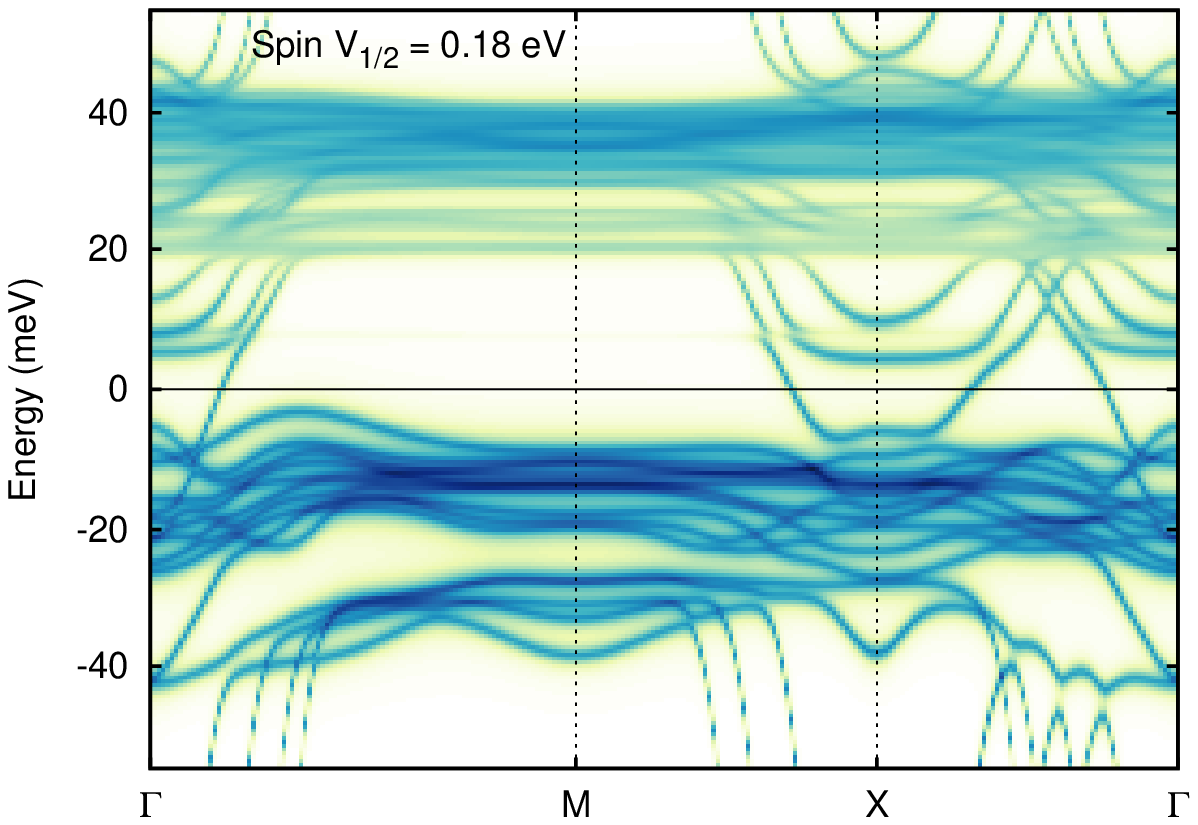}
\hspace{2mm}\includegraphics[width=1.15cm]{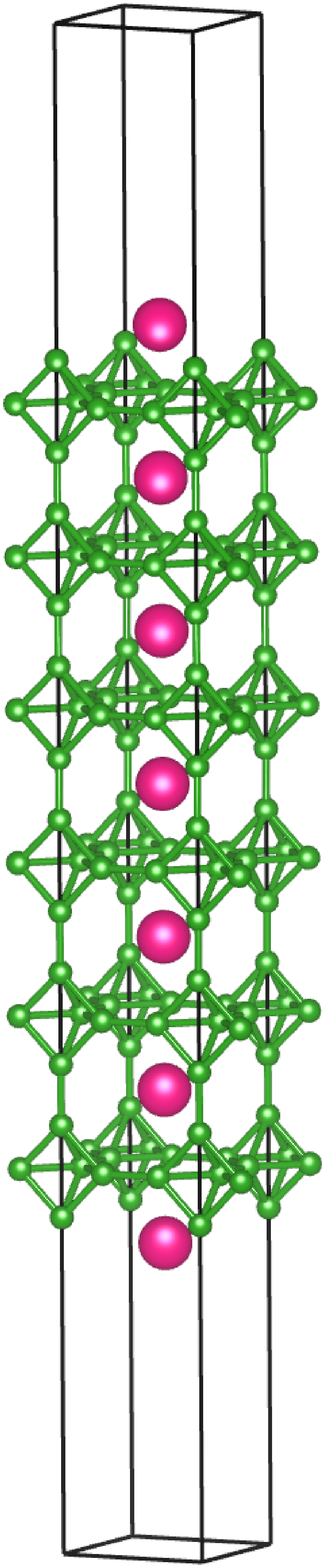}
\caption{
The $\mathbf{k}$-resolved DFT+DMFT spectrum (including all layers) of the Sm terminated SmB$_6$ supercell, shown in the right panel, in the presence of a spin-polarized ($B_z$) time-reversal symmetry breaking field. The left (middle) panel shows when the potential is applied to the sub-surface Sm 4f $J=5/2$, $j_z = \pm5/2$ ($j_z = \pm1/2$) states. The left panel shows a clear separation between the conduction and the valence bands: the perturbing field destroys the topological protection of the surface bands. In the middle panel, the field is applied to the $j_z = \pm1/2$ states, and a smaller gap emerges around the $X$-point, while the $\Gamma$-pocket remains ungapped.
\label{fig:slab-bands-pol}}
\end{figure*}

To numerically confirm the topological protection of the surface bands we applied time-reversal symmetric potentials to the (sub)surface layer of the Sm terminated supercell displayed to the right in Fig.~\ref{fig:slab-bands-pol}. The topological surface states should survive under these perturbing potentials, which can occur for example due to surface reconstructions. Our results are shown in Figs.~{\ref{fig:slabpotg}} and {\ref{fig:slabpotx}}, where we applied the potential to the
$j_z=\pm 5/2$ and $j_z=\pm 1/2$ spin-orbitals of the $J=5/2$ Sm $4f$ manifold, respectively. These spin-orbitals were chosen as the surface states around the $\Gamma$-point has primarily $j_z=\pm 5/2$ character, while the surface states around the $X$-point have mainly $j_z=\pm 1/2$ character. Technically, we simply added the potential term to the self-energy of these states after DMFT convergence (a self-consistency including this potential is beyond our illustrative purposes).
Both figures show the Sm $4f$ projection on the sub-surface (left panel) and subsub-surface states (right panel). Please remember that, as discussed in the main text, the surface layer itself has another Sm $4f$ valence and is insulating. The topological surface states hence appear in the sub-surface layer in the unperturbed systems.

Fig.~{\ref{fig:slabpotg}} (left) clearly shows that one of the flat $4f$ orbitals (i.e., the $j_z=\pm 5/2$)
 is shifted above the Fermi energy upon increasing the $j_z=\pm 5/2$ potential $V_{5/2}$. At the same time, the topological surface states around the $\Gamma$-point shift from the sub-surface layer for $V_{5/2}=0$ to the subsub-surface layer at $V_{5/2}=0.73\,$eV. There is a crossover in-between with the surface states being extended to both layers.

At the same time the topological surface states around the $X$-point remain on the sub-surface layer. If we instead apply a potential to the $j_z=\pm 1/2$ spin-orbitals as in Fig.~{\ref{fig:slabpotx}}, it is these topological surface states which start to shift to the next layer below. The $X$-pocket is however more robustly anchored to the sub-surface atom compared to the $\Gamma$-pocket, and does not shift completely away even for $V = 0.73\,$eV.

The  larger "mobility" of the $\Gamma$ pocket in the $j_z=\pm 5/2$ $J=5/2$ spin-orbitals is interesting, as it is seen in some experiments \cite{hlawenka2018trivial,Jiang2013,Xu13} but not in others \cite{Denlinger2013,Jiang2013,Xu13}. In the latter experiments, the $\Gamma$ pocket might simply hide a few layers deeper in the bulk, escaping its detection in surface-sensitive photoemission experiments. 

Another possible explanation of the experimental discrepancies reveals itself at $V_{5/2} = 0.36\,$eV in Fig.~\ref{fig:slabpotg}. At this potential strength the topologically derived $X$-pocket and $\Gamma$-pocket get very close to each other around the Fermi energy, and the latter can easily be misidentified as an umklapp-state \cite{hlawenka2018trivial,Xu13}. Interestingly, at the very same potential there are trivial bands which grace the Fermi energy close to the $\Gamma$-point. These trivial bands may be connected to the weak $\alpha$-pocket seen in some experiments\cite{hlawenka2018trivial,Jiang2013,Xu13}. However, to put these observations on firmer ground several fully self-consistent slab calculations with various realistic surface reconstructions are needed, which is beyond the scope of current study.


\subsection{Magnetic field perpendicular to the surface}
Next we apply a spin-polarizing field instead of a potential term to the surface. The field ($B_z$) is directed perpendicular to the surface and it breaks the time reversal symmetry of the system. The perturbation is hence expected to destroy the topological protection of the surface states.
We target the same two different pairs of states as for the potential term. That is, we apply the field to the subsurface Sm 4f $J=5/2$, $j_z = \pm5/2$ and $j_z = \pm1/2$ spin-orbitals. As shown in Fig.~\ref{fig:slab-bands-pol}, the perturbing fields indeed allow the surface states around the Fermi level to hybridize, including the otherwise protected topological surface states.

Let us start with the field applied to the $j_z = \pm5/2$ spin-orbitals in  Fig.~\ref{fig:slab-bands-pol} (left). 
The large coupling between the field and the $jz = \pm5/2$ component of the surface states makes the surface bands very susceptible to the perturbing (time reversal symmetry breaking) field. The topological surface states start to hybridize and an indirect band gap opens. Quantitatively, the spectrum is altered more at the $\Gamma$ point than at the $X$-point as here the $j_z = \pm5/2$ character dominate. But also at the $X$-point a small gap opens.

It is exactly vice versa if we apply the perturbing field to the subsurface Sm 4f $J = 5/2$, $j_z = \pm1/2$ spin-orbitals, see  Fig.~\ref{fig:slab-bands-pol} (right). Here, actually only the topological surface state around the $X$-point are gapped out, whereas the topological surface bands around $\Gamma$ remains intact. While applying a field to only part of the Sm $4f$ states is, as a matter of course, a theoretical construct, it still demonstrates that the $\Gamma$ and $X$ pockets can clearly shift independently. 

\end{document}